\documentclass[twocolumn,aps,prc,showpacs,floatfix]{revtex4}
\usepackage{graphicx}

\newcommand{\reac}{\bar{p}p\rightarrow\bar{\Lambda}\Lambda}
\newcommand{\fullreac}{\bar{p}p\rightarrow\bar{\Lambda}\Lambda
\rightarrow\bar{p}\pi^{+}p\pi^{-}}
\newcommand{\lbar}{\bar{\Lambda}}
\newcommand{\lbarl}{\bar{\Lambda}\Lambda}
\newcommand{\pbarp}{\bar{p}p}
\newcommand{\onehalf}{\frac{1}{2}}
\newcommand{\onefourth}{\frac{1}{4}}
\newcommand{\onefourpi}{\frac{1}{4\pi}}
\newcommand{\idt}{\mathcal{I}}
\newcommand{\Pvec}{\vec{P}}
\newcommand{\vsig}{\vec{\sigma}}
\newcommand{\lh}{\mathbf{\hat{l}}}
\newcommand{\mh}{\mathbf{\hat{m}}}
\newcommand{\nh}{\mathbf{\hat{n}}}
\newcommand{\al}{\alpha}
\newcommand{\alb}{\bar{\alpha}}
\newcommand{\alba}{\bar{\alpha}\alpha}
\newcommand{\abap}{\bar{\alpha}\alpha P^{T}}

\newcommand{\pt}{P^{T}}
\newcommand{\calL}{\mathcal{L}}
\newcommand{\pGeV} {\,\mbox{GeV}/\mbox{c}}
\newenvironment{bfrac}[2]
{\,\begin{array}{c} #1 \\ \hline #2 \end{array}\,}{}

\newcommand{\bbq}{^{\bar{B}}}
\newcommand{\pbar}{\bar{p}}

\newcommand{\kpom}{k^{\pbar}_{m}}
\newcommand{\kpon}{k^{\pbar}_{n}}
\newcommand{\kpol}{k^{\pbar}_{l}}
\newcommand{\kptm}{k^p_{m}}
\newcommand{\kptn}{k^p_{n}}
\newcommand{\kptl}{k^p_{l}}
\newcommand{\cphcm}{\cos{(\Phi_{c.m.})}}
\newcommand{\sphcm}{\sin{(\Phi_{c.m.})}}

\begin{document}

\title{Experimental determination of the complete spin structure
for\\  $\bar{p}p\rightarrow\bar{\Lambda}\Lambda$ at $p_{\bar{p}}= 1.637$ GeV/c}


\author{K.D.~Paschke}
\author{B.~Quinn}
\author{A.~Berdoz}
\author{G.B.~Franklin}
\author{P.~Khaustov}
\author{C.A.~Meyer}
\affiliation{Carnegie Mellon University,  Pittsburgh, Pennsylvania 15213}
\author{C.~Bradtke}
\author{R.~Gehring}
\author{S.~Goertz}
\author{J.~Harmsen}
\author{A.~Meier}
\author{W.~Meyer}
\author{E.~Radtke}
\author{G.~Reicherz}
\affiliation{Ruhr-Universit\"{a}t Bochum, D-44780 Bochum, Germany}
\author{H.~Dutz}
\author{M.~Pl\"{u}ckthun}
\author{B.~Schoch}
\affiliation{Universit\"{a}t Bonn, D-53115 Bonn, Germany}
\author{H.~Dennert}
\author{W.~Eyrich}
\author{J.~Hauffe}
\author{A.~Metzger}
\author{M.~Moosburger}
\author{F.~Stinzing}
\author{St.~Wirth}
\affiliation{Universit\"{a}t Erlangen-N\"{u}rnberg, D-91058 Erlangen, Germany}
\author{H.~Fischer}
\author{J.~Franz}
\author{F.H.~Heinsius}
\author{E.~Kriegler}
\author{H.~Schmitt}
\affiliation{Universit\"{a}t Freiburg, D-79104 Freiburg, Germany}
\author{B.~Bunker}
\author{D.~Hertzog}
\author{T.~Jones}
\author{R.~Tayloe}
\altaffiliation[Now at ]{Indiana University,Bloomington, IN, 47405}
\affiliation{University of Illinois,Urbana, Illinois 61801}
\author{R.~Br\"{o}ders}
\author{R.~Geyer}
\author{K.~Kilian}
\author{W.~Oelert}
\author{K.~R\"{o}hrich}
\altaffiliation[Now at ]{Creative Services, St. Genis, France}
\author{K.~Sachs}
\author{T.~Sefzick}
\affiliation{Institut f\"{u}r Kernphysik des Forschungszentrums J\"{u}lich, D-52425 J\"{u}lich, Germany}
\author{B.~Bassalleck}
\author{S.~Eilerts}
\author{D.E.~Fields}
\author{P.~Kingsberry}
\author{J.~Lowe}
\author{R.~Stotzer}
\affiliation{University of New Mexico, Albuquerque, New Mexico 87131}
\author{T.~Johansson}
\author{S.~Pomp}
\affiliation{Uppsala University, S-75121 Uppsala, Sweden}

\date{\today}

\begin{abstract}
The reaction $\bar{p}p\rightarrow\bar{\Lambda}\Lambda
\rightarrow\bar{p}\pi^{+}p\pi^{-}$ has been measured with 
high statistics at a beam
momentum of $p_{\bar{p}}=1.637$ GeV/c.  The use of a
transversely-polarized frozen-spin target combined with the
self-analyzing property of $\Lambda/\lbar$ decay allows access to
unprecedented information on the spin structure of the interaction.
The most general spin-scattering matrix can be written in terms of eleven
real parameters for each bin of scattering angle, each of these parameters is
determined with reasonable precision.
From these results all conceivable spin-correlations are
determined with inherent self-consistency.  
Good agreement is found with the few
previously existing measurements of spin observables 
in $\bar{p}p\rightarrow\bar{\Lambda}\Lambda$ near
this energy.  Existing theoretical models
do not give good predictions for those spin-observables that had not
been previously measured.
\end{abstract}


\maketitle

\section{Introduction}

Measurements have been made of the $\reac$ reaction on a transversely
polarized frozen-spin target.  An analysis is presented of data taken
at the Low Energy Antiproton Ring (LEAR) at CERN, Geneva at $p_{\pbar}= 1.637$ GeV/c
corresponding to a center-of-mass energy which is 78 MeV above
$\lbarl$ threshold. 

This experiment, carried out by the PS185
collaboration, expands upon a series of $\bar{Y}Y$ production and
related experiments \cite{old_PS185,Horsti,threshold,recent_PS185} 
which have previously been performed by this
collaboration using the same detector system.  These covered wide-ranging
kinematics from very near threshold to higher energies 
at which the larger cross section allowed high-statistics studies to be made.

Spin observables have long been of interest in strangeness-production reactions in
part because of the characteristic strong polarization produced.  The
hyperons in the final state (the word 'hyperon' will be used loosely
to include anti-hyperons) lend themselves to study of spin dynamics
because of the self-analyzing power of mesonic hyperon decay.  Thus,
even before the introduction of initial-state polarization in these
studies, the PS185 collaboration has
been able to greatly expand the world's supply of data on spin
observables, especially for $\reac$.  In particular, final-state 
polarization components of both
hyperons and correlations of all spin components of
$\Lambda$ and $\lbar$ were accurately measured in addition to
differential cross sections.

The wealth of data produced by PS185
\cite{old_PS185,Horsti,threshold,recent_PS185}
excited a great deal of theoretical activity
\cite{MEX_old,MEX_QG,MEX_pred,QG_old,QG_pred,other_theory} with several models enjoying 
reasonable success in fitting the observations.
Two distinct theoretical approaches, meson-exchange [MEX] and
quark-gluon [QG]
inspired models, have been used successfully to fit the
$\reac$ data despite being based on fundamentally different reaction
dynamics.  

Several authors \cite{MEX_old,MEX_QG,MEX_pred} have constructed models based
on $t$-channel exchange of strange mesons.  In order to match the
observed data, these meson-exchange 
models require a strong tensor interaction and so a
spin-flip from the initial spin-triplet $\pbarp$ pair to the final
spin-triplet $\lbarl$ pair.  Initial- and final-state interactions are
modeled as well but are found not to qualitatively change the
spin-flip character of these models.

The alternate quark-gluon inspired models\cite{QG_old,QG_pred} are based upon an
assumed $s$-channel interaction between $\bar{q}q$ pairs leading to the
transformation to an $\bar{s}s$ pair.  In existing models the
$\bar{q}q$ pair is assumed to have vacuum quantum numbers ($0^+$
for $^3P_0~\bar{q}q$ pairs) or gluon quantum numbers ($1^-$ for 
$^3S_1~\bar{q}q$ pairs).  Again there is a dominance of the
spin-triplet state (inspired by the empirical fact that the
singlet fraction is, on average, small).  In fact this construction assures pure triplet
transitions.  (Since the spin of the $\Lambda$ reflects the spin of the
$s$-quark, in the constituent quark model, triplet  $\bar{s}s$ 
pairs guarantee a triplet $\lbarl$ final
state.)   Here the assumed fundamental $s$-channel
$\bar{q}q\rightarrow\bar{s}s$ process does not flip the spin at all
but some spin-flip strength is introduced with the inclusion of
initial- and final-state interactions.

The difference in spin-flip predictions of MEX and QG models 
was suggested \cite{Holinde}
as a means of distinguishing the validity of the two classes of
models.  
Since each model predicted strong triplet
interactions, final-state spin correlations were similar and  
measurement of final-state spin alone was not sufficient to
distinguish them.  
Sensitivity
could be found in measurement of the correlation between initial-state
spin and final-state spin.  In particular, the normal-to-normal
depolarization and spin transfer were expected to strongly select
between the two classes of models.  These observables, often denoted
$D_{nn}$ and $K_{nn}$, respectively, are sensitive to the transfer of
spin from the initial-state proton to the final-state $\Lambda$ (in the
case of $D_{nn}$) or $\lbar$ (in the case of $K_{nn}$).  The 'n'
subscripts indicate that the component of spin considered in each case
is the component normal to the scattering plane.  Since a large
number of spin correlations will be discussed here, a more general
notation will be introduced below.  In particular,  $D_{nn}$ and $K_{nn}$
will be denoted as $Q[n_p,n_\Lambda]$ and $Q[n_p,n_{\lbar}]$,
respectively.

Experimental study of these particularly interesting observables
required a polarized target.
 A frozen-spin target was constructed
with such small dimensions that $\lbarl$ pairs could exit the target
before decaying.  The success of the present measurements relied upon
 the PS185 apparatus, this new target, and the superb
properties of the LEAR beam.

It was also noted \cite{Kent_Brian} that such data contained enough
information to permit determination of not just a few observables but
the entire spin-structure of the reaction.  From this, all possible
spin observables could then be determined.  That analysis has been
successfully carried out \cite{Kent_thesis} and is reported here.  
A previous publication
\cite{Bernd_et_al} already reported the measured spin-transfer and
depolarizations, extracted using the techniques which are explained in
the present publication.  Interestingly, the measured values disagree
strongly with the predictions from both classes of models, leaving
opportunity for further theoretical study.  These results are also
included in the present paper.

\section{Spin Correlations}

The density-matrix formalism lends itself to analysis of systems 
composed of ensembles of non-interfering states, such as occur in
polarized systems.   This formalism will be used therefore to
precisely define the spin-correlations and to relate them to the
observed final-state distribution on the one hand and to the
spin-scattering matrix, which parameterizes the transition, on the other.

In this formalism, the expectation value of an observable represented
by operator $O$ can be written as $<O>=Tr(O\rho^{\lbarl})$ where
$\rho^{\lbarl}$ is a $4 \times 4$ density matrix representing the
final-state spin information.  In Sec.~\ref{spincor1}, we write
$\rho^{\lbarl}$ in terms of the initial-state polarization vectors, 
$\vec{P}^{\bar{p}}$ and $\vec{P}^p$, and the $\reac$ transition
operator $M(\Theta_{c.m.})$.  Since the experiment actually detects the
final-state protons, antiprotons, and pions from the hyperon decays, the
formalism is extended in Sec.~\ref{spincor2} to relate the directions of
these particles' momenta to the initial-state polarization vectors and
the complete set of the reaction's spin-observables.

\subsection{Spin Dynamics of $\reac$}
\label{spincor1}
The initial-state density matrix for spin-$\frac{1}{2}$ particles with 
polarization $\vec{P}$ can be written as
\[\rho = \onehalf \left( \idt + \Pvec\cdot\vsig \right) \]
where $\idt$ is the identity matrix and $\sigma_i$ are the Pauli matrices.
In the following discussion, notation is greatly simplified by defining 
$P_0=1$ and $\sigma_0=\idt$.  Then
\[\rho = \onehalf \sum_{k=0}^{3}P_{k}\sigma_{k}  .\]
For an initial state of a proton and anti-proton the initial-state density 
matrix is
\begin{equation}
\label{eqn1}
\rho^{\pbarp} =  \rho^{\bar{p}} \rho^{p} 
= \onefourth  \sum_{j,k=0}^{3}P_{j}^{\bar{p}} \sigma_{j}^{\bar{B}} 
                                      P_{k}^{p}\sigma_{k}^{B}
\end{equation}
where $\sigma_{j}^{\bar{B}}$ and $\sigma_{k}^{B}$ operate in the separate 
spin-space of the anti-proton and proton, respectively, and a direct product 
is implied.  

Meanwhile, the density matrix after the interaction, $\rho^{\lbarl}$, can be 
written as an arbitrary linear combination of direct products of  
$\sigma_{\mu}^{\bar{B}}$ and $\sigma_{\nu}^{B}$ since these span the 
space of Hermitian 4$\times$4 operators
\begin{equation}
\label{eqn2}
\rho^{\lbarl}=  \sum_{\mu,\nu=0}^{3} W_{\mu\nu} \sigma_{\mu}^{\bar{B}} 
                                     \sigma_{\nu}^{B}
\end{equation}
where the notation of $\sigma^{\bar{B}}$ and $\sigma^{B}$ from Eq.~(\ref{eqn1}) have
been used in anticipation of the fact that an identification will be made
between the $\bar{p}$ and $\bar{\Lambda}$ anti-baryon spin spaces and 
between the $p$ and $\Lambda$ baryon spin spaces.

Spin correlations are meaningful observables as long as the 
coordinate system associated with each particle is well defined.
It is not necessary that the coordinate axes used for one particle be aligned
parallel to those used for another.
The following discussion of spin correlations applies for any set of 
coordinates in which the spin components of each particle are measured 
in some set of axes defined in the rest frame of that particle.
The specific choice of coordinate axes for this analysis is described in 
Sec.~\ref{spincor3} below.

It then follows (since $Tr(\sigma_\mu \sigma_\nu)=2\delta_{\mu\nu}$) that
\[4 W_{\mu\nu} = Tr( \sigma_{\mu}^{\bar{B}} \sigma_{\nu}^{B} \rho^{\lbarl})  .\]
Substituting this into Eq.~(\ref{eqn2}) gives
\begin{equation}
\label{eqn3}
\rho^{\lbarl}=  \frac{1}{4} \sum_{\mu,\nu=0}^{3}
               Tr( \sigma_{\mu}^{\bar{B}} \sigma_{\nu}^{B} \rho^{\lbarl}) 
                \sigma_{\mu}^{\bar{B}} \sigma_{\nu}^{B}  .
\end{equation}
On the right hand side $\rho^{\lbarl}$ can be rewritten in 
terms of $\rho^{\pbarp}$,
which is given in Eq.~(\ref{eqn1}).  If M is the transition operator from the
$\pbarp$ to $\lbarl$ state,
\begin{equation}
\label{eqn_a}
\rho^{\lbarl}= M \rho^{\pbarp} M^\dagger = 
\onefourth \sum_{j,k=0}^{3}P_{j}^{\bar{p}} P_{k}^{p} M \sigma_{j}^{\bar{B}} 
                                      \sigma_{k}^{B} M^\dagger  .
\end{equation}
Substituting this in Eq.~(\ref{eqn3}) gives the identity
\begin{equation}
\label{eqn4}
\rho^{\lbarl}=  \onefourth \sum_{\mu,\nu=0}^{3} \sum_{j,k=0}^{3}
P_{j}^{\bar{p}} P_{k}^{p} \frac{1}{4}
        Tr( \sigma_{\mu}^{\bar{B}} \sigma_{\nu}^{B} 
             M \sigma_{j}^{\bar{B}} \sigma_{k}^{B} M^\dagger)
                \sigma_{\mu}^{\bar{B}} \sigma_{\nu}^{B}  .
\end{equation}

Only the first term contributes to the unpolarized differential cross section,
\[I_0 = \frac{Tr(\rho^{\lbarl})}{Tr(\rho^{\pbarp})}= \frac{1}{4} Tr(M M^\dagger).\]
Factoring this out of the sum in Eq.~(\ref{eqn4}) and defining
\begin{equation}
\label{eqnQ}
Q[j_{\bar{p}},k_p,\mu_{\bar{\Lambda}},\nu_\Lambda] =
      \frac{1}{4 I_0} Tr( \sigma_{\mu}^{\bar{B}} \sigma_{\nu}^B 
             M \sigma_{j}^{\bar{B}} \sigma_{k}^B M^\dagger)
\end{equation}
allows Eq.~(\ref{eqn4}) to be written as
\begin{equation}
\label{eqn5}
\rho^{\lbarl}=  \onefourth I_0 \sum_{\mu,\nu=0}^{3} \sum_{j,k=0}^{3}
P_{j}^{\bar{p}} P_{k}^{p} 
Q[j_{\bar{p}},k_p,\mu_{\bar{\Lambda}},\nu_\Lambda]
                \sigma_{\mu}^{\bar{B}} \sigma_{\nu}^B  .
\end{equation}

The spin dynamics of the $\reac$ reaction at any production angle, 
$\Theta_{c.m.}$, is entirely contained within the quantities 
$Q[j_{\bar{p}},k_p,\mu_{\bar{\Lambda}},\nu_\Lambda]$.  These $Q$'s are
the observables of the present experiment, which we shall refer 
to as the 'spin correlations'.  The unpolarized differential cross 
section is parameterized by $I_0$.  Although the functional dependence 
will not be written explicitly, it is to be understood that $I_0$ and 
the $Q$'s are functions 
of $\Theta_{c.m.}$.  The Q's would be directly measurable if final-state 
spins could be measured directly (and initial-state polarizations 
could be chosen arbitrarily).
For example, measurement of the mean value of the product of the $\eta$-component of 
$\bar{\Lambda}$ spin and the $\xi$-component of $\Lambda$ spin (given the 
initial $\bar{p}$ polarized in the r direction and proton polarized in the 
s direction)
would be directly related to
$Q[r_{\bar{p}},s_p,\eta_{\bar{\Lambda}},\xi_\Lambda]$ by
\[<\sigma_{\eta}^{\bar{B}} \sigma_{\xi}^B> = 
    \frac{Tr(\rho^{\lbarl} \sigma_{\eta}^{\bar{B}} \sigma_{\xi}^B)}
                 {Tr(\rho^{\lbarl})}=
    \frac{P_{r}^{\bar{p}} P_{s}^{p} 
      Q[r_{\bar{p}},s_p,\eta_{\bar{\Lambda}},\xi_\Lambda]}{\mathcal{N}}\]
where the normalization factor is given by
\begin{eqnarray*}
     \mathcal{N}= 1&+&P_{r}^{\bar{p}}Q[r_{\bar{p}},0,0,0]
 + P_{s}^{p}Q[0,s_p,0,0]\\
      &+ &P_{r}^{\bar{p}} P_{s}^{p}Q[r_{\bar{p}},s_p,0,0]
\end{eqnarray*}

The redundant particle-identification subscripts have been introduced 
in the index list of Q to allow suppression of vanishing elements.
Subsequently terms such as $Q[0,s_p,0,0]$ will be written simply as 
$Q[s_p]$.

\subsection{Angular Distribution of Decay Products}
\label{spincor2}

In fact $\lbarl$ final-state spin information is not directly measured in this 
experiment.  It can be inferred, however, from the angular distribution of the
decay products because of the self-analyzing nature of the $\Lambda$
and $\bar{\Lambda}$ decays.  For a $\Lambda$ with polarization vector 
$\vec{P}^\Lambda$, the angular distribution of the decay proton in the $\Lambda$
rest frame is given by
\[I_p(\hat{k}^p) = \onefourpi (1 + \alpha \vec{P}^\Lambda \cdot \hat{k}^p)\]
where $\alpha = 0.642 \pm .013$ is the self-analyzing power \cite{PDG} of $\Lambda
\rightarrow p\pi^{-}$ and $\hat{k}^p$ is a unit vector in the direction of the 
proton's momentum (in the $\Lambda$ rest frame).
Similarly, for $\bar{\Lambda}\rightarrow\bar{p}\pi^{+}$,
\[I_{\bar{p}}(\hat{k}^{\bar{p}}) = \onefourpi (1 + \bar{\alpha} 
\vec{P}^{\bar{\Lambda}} \cdot \hat{k}^{\bar{p}})\]
where $\bar{\alpha} = -\alpha$ by CP-conservation.

The transition operator $T_\Lambda$, representing
$\Lambda \rightarrow p\pi^{-}$ decay must give this observed angular
distribution from
\[I_p(\hat{k}^p) = Tr(T_\Lambda \rho^\Lambda T_\Lambda^\dagger)
   = Tr(T_\Lambda \frac{1}{2}(\idt+\vec{P}^\Lambda\cdot\vec{\sigma}) T_\Lambda^\dagger)\]
It then follows that
\begin{equation}
\label{eqn6}
Tr(T_\Lambda \sigma_i T_\Lambda^\dagger) = \frac{\alpha k_i^p}{2 \pi}
\text{\ \ \ \ if $i$ }\epsilon \{1,2,3\}
\end{equation}
(where  $k_i^p$ is the $i$'th directional cosine of the proton's momentum) while
\begin{equation}
\label{eqn7}
Tr(T_\Lambda \sigma_i T_\Lambda^\dagger) = Tr(T_\Lambda T_\Lambda^\dagger) = \frac{1}{2 \pi}\text{\ \ \ \ if $i$=0}
\end{equation}
For notational simplification, Eqs.~(\ref{eqn6}) and (\ref{eqn7}) can be 
combined into a single equation of the form of Eq.~(\ref{eqn6}) by extending
the definition of $k^p_\mu$, defining $k^p_0 = \frac{1}{\alpha}$.
Then Eq.~(\ref{eqn6}) can be used to find the angular distribution of the 
final-state proton and anti-proton resulting from the decay of the $\Lambda$
and $\bar{\Lambda}$,
\begin{widetext}
\begin{equation}
\label{eqn8}
I_{final}(\Theta_{c.m.},\Phi_{c.m.},\hat{k}^{\bar{p}},\hat{k}^p) =
    Tr(T_{\bar{\Lambda}} T_\Lambda \rho^{\lbarl} 
              T_{\bar{\Lambda}}^\dagger T_\Lambda^\dagger)
 =
\frac{I_0(\Theta_{c.m.})}{16\pi^2}
{\sum_{\mu,\nu=0}^{3} \sum_{j,k=0}^{3}
\bar{\alpha}\alpha 
Q[j_{\bar{p}},k_p,\mu_{\bar{\Lambda}},\nu_\Lambda]P_{j}^{\bar{p}} P_{k}^{p}
k^{\bar{p}}_\mu k^p_\nu}
\end{equation}
\end{widetext}
where $k^{\bar{p}}_\mu$ has similarly been extended by defining 
$k^{\bar{p}}_0 = \frac{1}{\bar{\alpha}}$.

Eq.~(\ref{eqn8}) provides the connection between the observed
distribution, $I_{final}$, and the spin correlations, 
$Q[j_{\bar{p}},k_p,\mu_{\bar{\Lambda}},\nu_\Lambda]$,
which are functions of the production angle $\Theta_{c.m.}$ and of the beam 
energy.  If no further constraints were known, the spin correlations could 
be extracted (for each bin of $\Theta_{c.m.}$) from the observed decay 
angular distributions.
In fact, as discussed in the next section, the structure of the spin-scattering
matrix is subject to constraints which enforce parity and charge
conjugation symmetry. As a result the spin 
correlations are not independent functions.
Rather than attempting to determine spin correlations independently, 
it is preferable to determine the parameters of the spin-scattering matrix
and to use them to determine the spin correlations of interest.  This results
in improved precision of the resulting spin correlations and guarantees that
they are mutually consistent, obeying all constraints implied by the
structure of the allowed spin-scattering matrix \cite{Richard}.  
It has been shown in reference 
\cite{Kent_Brian} that measurements of $\fullreac$ with an unpolarized
beam on a transversely polarized target provide enough information to fully 
constrain the spin-scattering matrix.  This then allows determination of
all spin correlations, even those with non-vanishing $j_{\bar{p}}$, for 
example, whose direct measurement would require the use of polarized beam.

\subsection{Coordinate systems}
\label{spincor3}

The emphasis of this experiment was the determination of correlations 
betweens spins, particularly between initial-state target spin and and 
final-state spins.
Comparison of initial- and final-state spin is complicated by the fact 
that initial-state particles are moving at relativistic velocity ($\beta$=.58)
in the center of mass.
Comparing all spins in a common reference frame would then require 
a relativistic boost of the spins.
Such a boost of 2-D spinors is not well defined.
Consistent transformation to a common reference frame would require use 
of relativistic 4-spinors.
This complication is avoided by defining all such correlations in terms of
the spin projections of each particle in its own rest frame.
As mentioned above, it is not necessary for the coordinate axes used
to  describe the spin of one particle be parallel to those
used to describe the spin of another.
Indeed, since there is no common boost direction between 
particle rest 
frames it is not generally possible to define the coordinate axes to be all
mutually parallel.  It is conventional to use helicity-based
coordinates in which one axis is aligned with the particle's
center-of-mass momentum direction.

The final-state polarization direction is inferred from the angular 
distributions of the decay products in the rest frames of the $\Lambda$ and
$\bar{\Lambda}$.  It is therefore natural to express spin information for each 
of these hyperons with respect to coordinate axes which are defined in their 
respective rest frames. 
Similarly, it is natural to express the target spin 
information with respect to axes which are at rest in the lab frame.  For 
notational completeness, although the beam is unpolarized, its spin is 
expressed in a coordinate system moving with the incident $\bar{p}$.

\begin{figure}
\includegraphics[width=8.6 cm,clip]{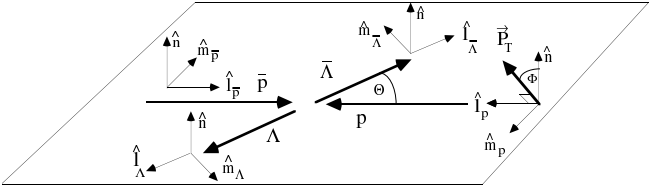}
\caption[]{\label{coordinates}
The coordinate axes used to decompose each of the individual
particles' spin directions are shown.  The $\hat{n}$ direction is
common.  The target polarization direction, which is perpendicular to
$\hat{l}_p$ is also shown for a typical
event in which the normal to the scattering plane is at angle $\Phi$
relative to the polarization direction.
}
\end{figure}

The specific coordinate systems used in this analysis are represented
in Fig.~\ref{coordinates}.  All four coordinate systems share a common $\nh$ 
direction, which is the normal to the scattering plane defined as a unit
vector in the direction of $\vec{\bf p}_p \times  \vec{\bf p}_{\Lambda}$ (or 
equivalently in the direction of $\vec{\bf p}_{\bar{p}} \times 
 \vec{\bf p}_{\bar{\Lambda}}$).  For the coordinate system associated with 
each particle a second axis ($\lh_p,\ \lh_{\bar{p}},\ \lh_{\Lambda},
\ \lh_{\lbar}$) is defined along the direction of the particle's 
momentum with respect to the center of mass of the $\pbarp$ system.  Finally 
the third axis ($\mh_p,\ \mh_{\bar{p}},\ \mh_{\Lambda},\ \mh_{\lbar}$) 
is defined for each coordinate system as $\mh_i = \nh_i \times \lh_i$ so each 
($\lh_i,\ \mh_i,\ \nh_i$) defines a right-handed orthonormal set of basis 
vectors in the rest frame of particle i.

It is of course important to take into account these coordinate definitions when the 
results presented here are compared to predictions or to other 
measurements.  The meaning of a specific spin correlation depends upon
the axes with respect to which it is defined.  For example, with this choice 
of coordinates, the angle between $\mh_p$ and $\mh_{\Lambda}$ depends
upon the scattering angle.  This must not be neglected when interpreting 
a result such as the correlation between the $\mh_p$-component of the 
initial proton spin and the $\mh_{\Lambda}$-component of the final 
$\Lambda$ spin.  
These coordinates were also used in extracting the previously 
published results \cite{Bernd_et_al} from this measurement.  Since
those results involved correlations between normal components
of  spin, there was no potential for ambiguity in interpretation of 
those results.

With these coordinate definitions, $\nh$ is an axial vector while $\mh$ 
and $\lh$ are polar vectors for each of the coordinate systems.  
Then for any of the spins
expressed in its respective coordinate system, $\sigma_n$ is scalar under 
parity inversion while $\sigma_l$ and $\sigma_m$ are pseudoscalar.  Parity
conservation then requires that 
$Q[r_{\bar{p}},s_p,\eta_{\bar{\Lambda}},\xi_\Lambda]$ must vanish
 if an odd number of l's and m's appear in 
\{r,s,$\eta$,$\xi$\}.
Since the target is transversely polarized in this experiment $P^p_l$ is zero.
This reduces the sum over $k$ in Eq.~(\ref{eqn8}) to run only over
$k$ $\epsilon$ \{0,$m$,$n$\}.  Furthermore, the unpolarized beam means that 
only the j=0 terms survive.  Further constraints on Eq.~(\ref{eqn8}) 
result from C-parity conservation which requires that
$Q[j_{\bar{p}},k_p,\mu_{\bar{\Lambda}},\nu_\Lambda] = 
Q[k_{\bar{p}},j_p,\nu_{\bar{\Lambda}},\mu_\Lambda]$.  Additional 
simplification results from Bohr's rule \cite{Bohr} which requires, for example
that 
$Q[n_p,m_{\bar{\Lambda}},m_\Lambda] = -Q[n_p,l_{\bar{\Lambda}},l_\Lambda]$.
Incorporating all these simplifications, Eq.~(\ref{eqn8}) reduces from
a sum over 256 terms to
\begin{widetext}
\begin{equation}\label{eqn_b}
\frac{I_{final}(\Theta_{c.m.},\Phi_{c.m.},\hat{k}^{\bar{p}},\hat{k}^p)}
{I_0(\Theta_{c.m.})/16\pi^{2} }
 =
\left[
\begin{array}{lll}
& \hspace{0.15cm} 1 \\
+ & Q[n_\Lambda] &  (\alb \kpon + \al \kptn ) \\
+ & Q[n_{\bar{\Lambda}},n_\Lambda] &  \alba \kpon \kptn \\
+ & Q[m_{\bar{\Lambda}},m_\Lambda] &  \alba \kpom \kptm \\
+ & Q[l_{\bar{\Lambda}},l_\Lambda] &  \alba \kpol \kptl \\
+ & Q[m_{\bar{\Lambda}},l_\Lambda] &  \alba ( \kpom \kptl + \kpol \kptm ) \\
+ & Q[n_p] & ( \pt \cphcm + \abap \kpon \kptn \cphcm )\\
+ & Q[n_p,n_{\bar{\Lambda}}] & \alb \pt \kpon \cphcm \\
+ & Q[n_p,n_\Lambda] &  \al \pt \kptn \cphcm \\
+ & Q[m_p,m_{\bar{\Lambda}}] & \alb \pt \kpom \sphcm \\
+ & Q[m_p,l_{\bar{\Lambda}}] & \alb \pt \kpol \sphcm \\
+ & Q[m_p,m_\Lambda] & \al \pt \kptm \sphcm \\
+ & Q[m_p,l_\Lambda] & \al \pt \kptl \sphcm \\
+ & Q[n_p,m_{\bar{\Lambda}},m_\Lambda] & \abap ( \kpom\kptm\cphcm - \kpol\kptl\cphcm ) \\
+ & Q[n_p,m_{\bar{\Lambda}},l_\Lambda] & \abap \kpom \kptl \cphcm \\
+ & Q[n_p,l_{\bar{\Lambda}},m_\Lambda] & \abap \kpol \kptm \cphcm \\
+ & Q[m_p,m_{\bar{\Lambda}},n_\Lambda] & \abap \kpom \kptn \sphcm \\
+ & Q[m_p,l_{\bar{\Lambda}},n_\Lambda] & \abap \kpol \kptn \sphcm \\
+ & Q[m_p,n_{\bar{\Lambda}},m_\Lambda] & \abap \kpon \kptm \sphcm \\
+ & Q[m_p,n_{\bar{\Lambda}},l_\Lambda] & \abap \kpon \kptl \sphcm 
\end{array}\right].
\end{equation}
\end{widetext}
Eq.~(\ref{eqn_b})  contains just 19 spin-correlations along with $I_0$.  Since each term
manifests a unique angular dependence these spin-correlations are, in principle,
directly measurable by fitting Eq.~(\ref{eqn_b}) to the angular distribution
observed in scattering from a transversely polarized target.  Although this is
not the technique employed here for determination of spin-correlations, these
19 spin-correlations will be referred to as the being ''directly measurable''.

\section{Spin-scattering Matrix}

The transition operator M, introduced in Eq.~(\ref{eqn_a}), transforms 
from the space spanned by direct products of proton and anti-proton 
spinors to one spanned by the direct products of $\Lambda$ and 
$\lbar$ spinors.  Since it includes the spin part of the 
transition, it is called the spin-scattering matrix.  It can be 
represented by a $4\times4$ complex matrix.  It is convenient to construct 
this operator from a combination of direct products of 'baryon operators' 
and 'anti-baryon operators'.   Baryon operators will be defined to be those 
which transform from the proton spin-space to the $\Lambda$ spin-space 
while leaving anti-baryon spinors unaffected.  Conversely, anti-baryon 
operators will transform from $\bar{p}$ spinors into $\lbar$ spinors.  
This identification of proton spinors with $\Lambda$ spinors is just a 
convenience, there is no loss of generality since the 16 direct 
products considered will span the entire space of $4\times4$ Hermitian
matrices.

Constructing terms with definite symmetry properties is simplified by 
choosing \{$\idt^B,\sigma_l^B,\sigma_m^B,\sigma_n^B$\} as the baryon 
operators and \{$\idt\bbq,\sigma_l\bbq,\sigma_m\bbq,\sigma_n\bbq$\} as the 
anti-baryon operators.
Although these operators have the same matrix representations as the
spin operators, they are actually transition matrices having the space
of proton spinors as their domain and the $\Lambda$ spinor space as
their range. 
For $\eta\ \epsilon$ \{0, $l$, $m$, $n$\}, each $\sigma^B_\eta$
transforms eigenstates of the 
$\mathbf{\hat{\eta}}$-component of proton spin to the same eigenstates of the 
$\mathbf{\hat{\eta}}$-component of $\Lambda$ spin, multiplied by the eigenvalue.  
Similarly, $\sigma_\eta\bbq$ maps anti-proton
$\mathbf{\hat{\eta}}$-eigenstates to anti-lambda
$\mathbf{\hat{\eta}}$-eigenstates.
Under parity 
inversion, the behavior of $\sigma_n^B$ differs from that of $\sigma_l^B$ 
and $\sigma_m^B$ because $\nh$ is an axial vector while $\lh$ and
$\mh$ are polar vectors.  Given that they act upon 
components of spinors and produce components of spinors, the $\sigma_\eta^B$'s 
have the 
same parity properties as the corresponding components of spin, 
i.e. $\sigma_n^B$ is scalar under parity inversion while $\sigma_l^B$ and 
$\sigma_m^B$ are pseudo-scalar.

Given these properties, the baryon and anti-baryon operators may be used 
to construct a complete set of operators having good parity and C-parity and 
spanning the direct-product space.  A set of such operators is listed in
Table \ref{tab_pc_operators} along with their parity and charge 
conjugation eigenvalues.

\begin{table}
\begin{tabular} {lccr}
\hline \hline
Operator~~~~~~~~~~~~ & ~~P~~ & ~~C~~ &~~coefficient\kill\\
\hline 
$\idt^B\idt\bbq$&+&+&$a+b$\\
$\sigma_n^B\sigma_n\bbq$&+&+&$a-b$\\
$\sigma_l^B\sigma_l\bbq$&+&+&$c+d$\\
$\sigma_m^B\sigma_m\bbq$&+&+&$c-d$\\
$\sigma_n^B\idt\bbq + \idt^B\sigma_n\bbq$&+&+&$e$\\
$\sigma_n^B\idt\bbq - \idt^B\sigma_n\bbq$&$+$&$-$&0\\
$\sigma_m^B\idt\bbq + \idt^B\sigma_m\bbq$&$-$&+&0\\
$\sigma_m^B\idt\bbq - \idt^B\sigma_m\bbq$&$-$&$-$&0\\
$\sigma_l^B\idt\bbq + \idt^B\sigma_l\bbq$&$-$&+&0\\
$\sigma_l^B\idt\bbq - \idt^B\sigma_l\bbq$&$-$&$-$&0\\
$\sigma_m^B\sigma_l\bbq + \sigma_l^B\sigma_m\bbq$&+&+&$g$\\
$\sigma_m^B\sigma_l\bbq - \sigma_l^B\sigma_m\bbq$&+&$-$&0\\
$\sigma_n^B\sigma_m\bbq + \sigma_m^B\sigma_n\bbq$&$-$&+&0\\
$\sigma_n^B\sigma_m\bbq - \sigma_m^B\sigma_n\bbq$&$-$&$-$&0\\
$\sigma_n^B\sigma_l\bbq + \sigma_l^B\sigma_n\bbq$&$-$&+&0\\
$\sigma_n^B\sigma_l\bbq - \sigma_l^B\sigma_n\bbq$&$-$&$-$&0\\
\hline \hline
\end{tabular}
\caption[dummy]{\label{tab_pc_operators}A complete set of operators 
spanning the space of 
operators which transform from $\pbarp$ spinors to $\lbarl$ spinors.  These
operators are constructed to have definite parity P and charge-conjugation 
parity C.  The P and C values are also listed.  The final column
gives the coefficient by which each term is weighted in forming M.}
\end{table}

Conservation of parity and C-parity in the $\reac$ reaction requires 
\cite{Bystricky} that only the terms with positive parity and C-parity 
contribute to the spin-scattering matrix.  An arbitrary linear combination 
of the allowed terms can be constructed by weighting each term by the 
coefficients given in Table \ref{tab_pc_operators}.  A conventional
parameterization for M 
\cite{Mformat} is
\begin{equation}
\label{eqn9}
M=
\begin{array}{ll}
\frac{1}{2}&\left\{(a+b)\idt^B\idt\bbq+(a-b)\sigma_n^B\sigma_n\bbq+(c+d)\sigma_m^B\sigma_m\bbq\right.\\
&\ +(c-d)\sigma_l^B\sigma_l\bbq +e (\sigma_n^B\idt\bbq+\idt^B\sigma_n\bbq)\\
&\left.\ +g(\sigma_m^B\sigma_l\bbq+\sigma_l^B\sigma_m\bbq)\right\}\\
\end{array}
\end{equation}

Since overall phase is unimportant, the six complex parameters \{a,b,c,d,e,g\}
can be represented by just 11 real parameters for each $\Theta_{c.m.}$.  
Specifically, parameter 'a' will be chosen to be real and non-negative 
while the other five parameters have real and imaginary parts.
These 
parameters may be determined by an unbinned 11-parameter simultaneous 
fit to the observed production and decay angles of 
reconstructed $\fullreac$ events, as described in Sec.~\ref{fit} below.
Once the parameters of the spin-scattering matrix have been determined, a
wealth of spin-correlations can be calculated by substituting the form
of $M$ given by Eq.~(\ref{eqn9}) into the definition of the spin
correlation, Eq.~(\ref{eqnQ}).  
For example, Table \ref{tab_spin_corr} gives the results, in terms of
 \{a,b,c,d,e,g\}, for  the 24 spin correlations which could  be directly 
measured with a transversely polarized target and unpolarized beam.
In fact, Eq.~(\ref{eqnQ}) can also be used to calculate other 
spin correlations which would be directly measurable only with 
longitudinal target polarization and/or polarized beam.

Table \ref{tab_spin_corr} also lists names which have traditionally been 
used to identify some of these spin correlations, A for scattering asymmetry, 
P for final-state
polarization, D for depolarization, K for spin transfer, and C for 
final-state spin correlations.  Here C is also used for 3-spin correlations 
between the target and final state.  Caution should be used in identifying 
these spin correlations with those in other publications having the same 
traditional name.  The precise meaning of spin correlations involving m 
and l depend critically upon the choice of coordinates.

\begin{table*}
\begin{tabular} {lll}
\hline \hline
Spin Correlation & Traditional&$I_0\  \times\ Q \ \ \ 
      (\ =\frac{1}{4} Tr( \sigma_{\mu}^{\bar{B}} \sigma_{\nu}^B 
             M \sigma_{j}^{\bar{B}} \sigma_{k}^B M^\dagger)\ \ )$
   \\
\hline
$Q[0_{\bar{p}},0_p,0_{\bar{\Lambda}},0_\Lambda]=1$  & &$I_0= \frac{1}{2} \{ |a|^{2} + |b|^{2} + |c|^{2} + |d|^{2} + 
|e|^{2} + |g|^{2} \} $\\
$Q[n_{\bar{\Lambda}},n_\Lambda]$ &$ C_{nn } $&$ \frac{1}{2} \{ |a|^{2} - |b|^{2} - |c|^{2} + |d|^{2} + 
|e|^{2} + |g|^{2} \} $\\
$Q[n_p,n_\Lambda]$ &$ D_{nn } $&$ \frac{1}{2} \{ |a|^{2} + |b|^{2} - |c|^{2} - |d|^{2} + 
|e|^{2} - |g|^{2} \} $\\
$Q[n_p,n_{\bar{\Lambda}}]$&$ K_{nn } $&$ \frac{1}{2} \{ |a|^{2} - |b|^{2} + |c|^{2} - |d|^{2} + 
|e|^{2} - |g|^{2} \} $\\
$Q[n_\Lambda]=Q[n_{\bar{\Lambda}}]$ 
              &$ P_{n } = \bar{P}_n $&$  Re(a^{\ast}e) - Im(d^{\ast}g) $\\
$Q[n_p]=Q[n_p,n_{\bar{\Lambda}},n_\Lambda]$ 
              &$ A_n=C_{nnn } $&$  Re(a^{\ast}e) + Im(d^{\ast}g) $\\
$Q[m_{\bar{\Lambda}},l_\Lambda]=Q[l_{\bar{\Lambda}},m_\Lambda]$ 
              &$ C_{ml }=C_{lm} $&$  Re(a^{\ast}g) + Im(d^{\ast}e) $\\
$Q[n_p,m_{\bar{\Lambda}},m_\Lambda]=-Q[n_p,l_{\bar{\Lambda}},l_\Lambda]~~~~$
              &$ C_{nmm }=-C_{nll}~~~~ $&$ Re(d^{\ast}e) + Im(a^{\ast}g)  $\\
$Q[m_{\bar{\Lambda}},m_\Lambda]$ 
              &$ C_{mm } $&$  Re( a^{\ast}d + b^{\ast}c) + Im(e^{\ast}g) $\\
$Q[l_{\bar{\Lambda}},l_\Lambda]$ 
               &$ C_{ll } $&$ Re(-a^{\ast}d + b^{\ast}c) - Im(e^{\ast}g) $\\
$Q[n_p,l_{\bar{\Lambda}},m_\Lambda]$
               &$ C_{nlm } $&$ Re(e^{\ast}g) + Im(-a^{\ast}d + b^{\ast}c) $\\
$Q[n_p,m_{\bar{\Lambda}},l_\Lambda]$   
               &$ C_{nml } $&$ Re(e^{\ast}g) + Im(-a^{\ast}d - b^{\ast}c) $\\
$Q[m_p,m_\Lambda]$  
               &$ D_{mm } $&$ Re(a^{\ast}b + c^{\ast}d) $\\
$Q[m_p,n_{\bar{\Lambda}},l_\Lambda]$ 
               &$ C_{mnl } $&$ Im(-a^{\ast}b + c^{\ast}d) $\\
$Q[m_p,m_{\bar{\Lambda}}]$
               &$ K_{mm } $&$ Re(a^{\ast}c + b^{\ast}d) $\\
$Q[m_p,l_{\bar{\Lambda}},n_\Lambda]$
               &$ C_{mln } $&$ Im(-a^{\ast}c + b^{\ast}d) $\\
$Q[m_p,n_{\bar{\Lambda}},m_\Lambda]$
               &$ C_{mnm } $&$ Re(b^{\ast}e) - Im(c^{\ast}g) $\\
$Q[m_p,l_\Lambda]$ 
               &$ D_{ml } $&$  Re(c^{\ast}g) + Im(b^{\ast}e) $\\
$Q[m_p,l_{\bar{\Lambda}}]$ 
               &$ K_{ml } $&$  Re(b^{\ast}g) + Im(c^{\ast}e) $\\
$Q[m_p,m_{\bar{\Lambda}},n_\Lambda]$
               &$ C_{mmn } $&$  Re(c^{\ast}e) - Im(b^{\ast}g) $\\
\hline \hline
\end{tabular}

\caption[dummy2]{\label{tab_spin_corr}The first column lists the spin 
correlations which are directly accessible with a transversely polarized 
target and unpolarized beam.  The third column gives the indicated 
correlation (multiplied by $I_0$) in terms of the parameters of the 
spin-scattering matrix given in Eq.~(\ref{eqn9}).  The second column 
lists alternate names for the spin correlations.}
\end{table*}

\section{Apparatus}

\begin{figure*}
\includegraphics[width=8. cm,clip]{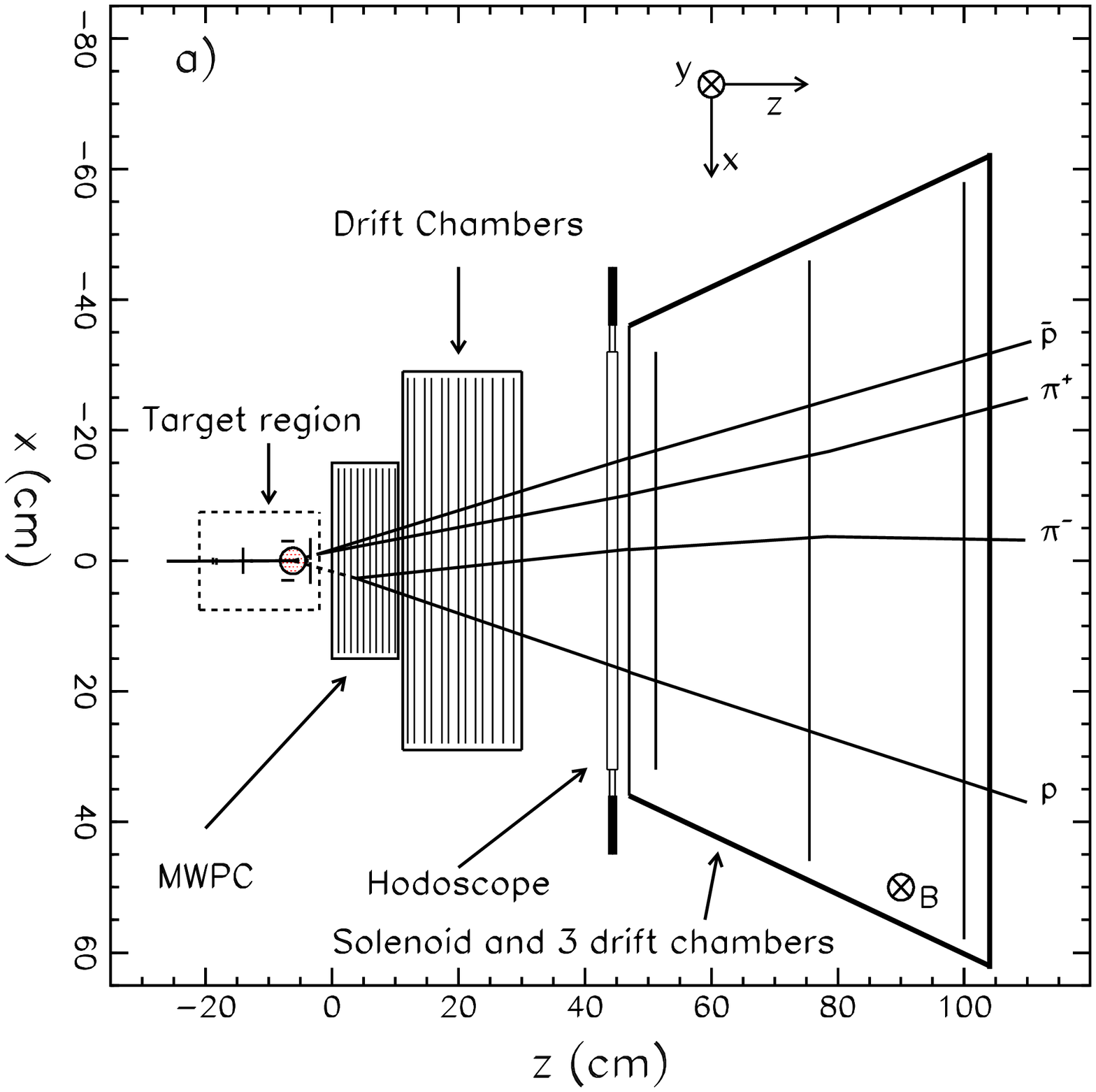}
\includegraphics[width=8. cm,clip]{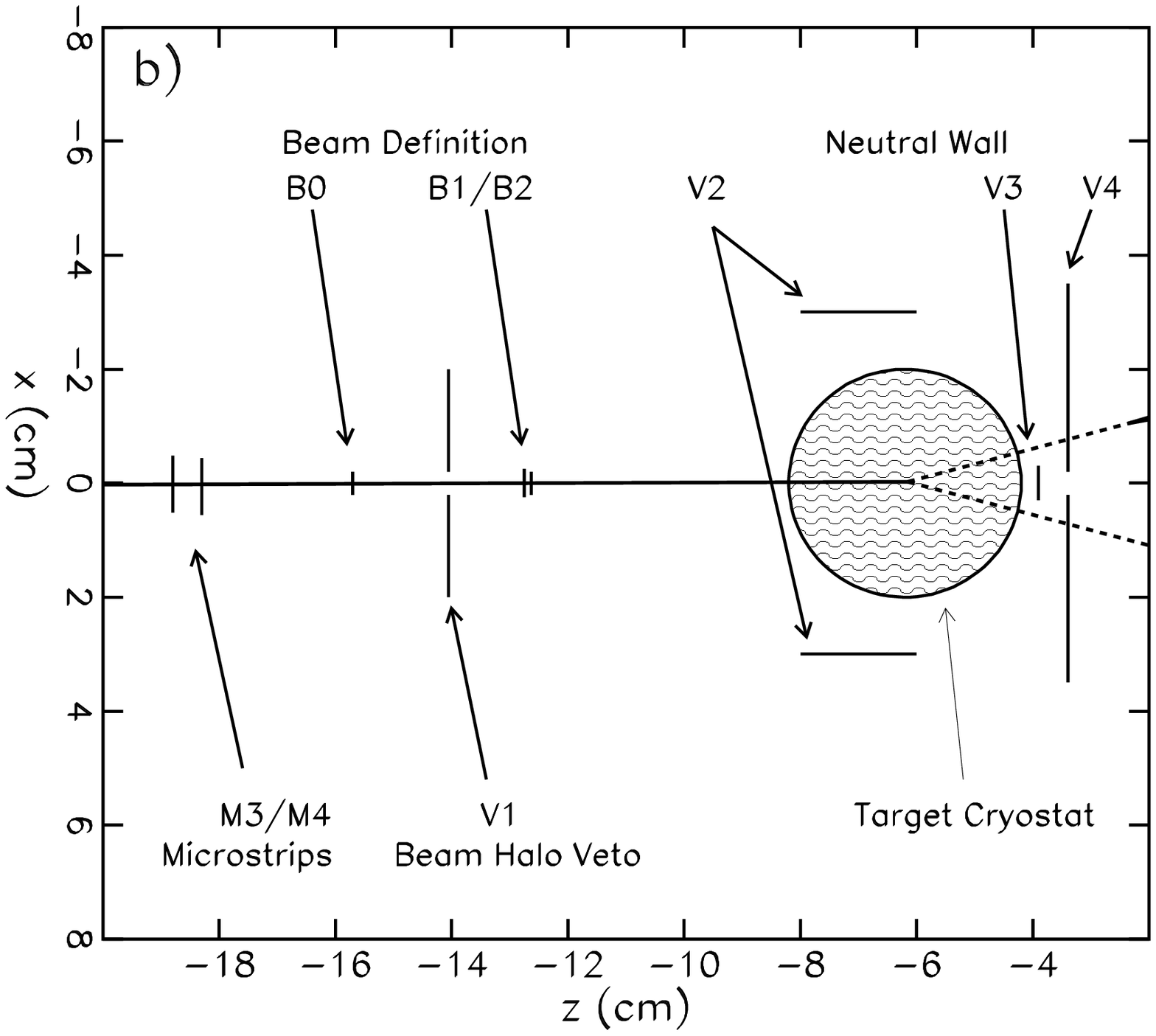}

\caption[]{
\label{detector_schematic}
a) A schematic plan view of the detectors, magnet, and target.
The solenoid field, shown
in the positive y-direction, could be reversed to reduce systematic uncertainties.
Tracks from a typical event are superimposed.
b) An expanded view of the target region shows the beam-defining 
scintillators and the
veto scintillators used to select events in which only neutral
particles exit the target.  The hodoscope planes completed the trigger 
by requiring charged particles downstream of the decay region.
}
\end{figure*}

With the exception of the polarized target and associated trigger detectors, 
most of the equipment was the same as that used in previous versions of the 
PS185 experiment
\cite{old_PS185,Horsti,threshold,recent_PS185}.
 A schematic view of the apparatus 
is shown in Fig.~\ref{detector_schematic}.  A compact set of tracking 
detectors was located just downstream of the target area.  
Because of the forward boost of 
the $\lbarl$ system, the trajectories of both particles passed through these 
tracking chambers.  This resulted in a large acceptance for the 
full reconstruction 
of the charged tracks resulting from $\fullreac$.  Accurate measurement of the
topology of these charged tracks is sufficient, apart from ambiguity of 
$\Lambda$ vs. $\lbar$, to completely reconstruct the kinematics of each event 
including the hyperon decay angles whose distributions yield 
information on the final-state spin.  The ambiguity of $\Lambda$ vs. $\lbar$ is 
resolved by a solenoid magnet further downstream which bends the trajectories 
enough to allow the sign of the charge of tracks to be determined.

The tracking detectors consisted of two detector stacks.  The first was 10 
planes of multi-wire proportional chambers with planes alternately oriented at 
$\pm 45^\circ$ relative to the horizontal (the U and V directions).
These had a pitch of 1.27 mm 
and a separation of 1 cm between planes.  The second stack was 13 planes of 
drift chambers oriented vertically and horizontally (the X and Y
directions) 
with an average separation 
between planes of 1.35 cm.  The 4 cm drift cells were read out by a pair of 
sense wires separated by .42 mm, which resolved the usual left-right 
ambiguity.  A set of three similar but larger drift chambers was 
located inside the solenoid magnet to determine the direction of 
deflection of tracks by the 1 kG magnetic field.  

Since the beam passed through all these chambers, it was necessary to 
desensitize the center of each chamber.  This was done by
electro-plating additional
metal onto a 3 to 10 mm length of the relevant sense wires to thicken 
them at the position at which the beam would pass, preventing gas 
amplification.

Four planes of microstrips with 100 $\mu$m pitch were located upstream 
of the target.  These provided tracking of the individual incident $\bar{p}$
for a fraction of the events.  
Four planes provided no redundancy so a beam track could be
reconstructed only for events in which exactly one cluster was found
on each plane.  This applied for only about 45\% of events because of
beam pile-up and detector-aging in the high intensity beam.
For other events the average beam track was used.

The target was enclosed within a 4.2 cm diameter vacuum vessel.   
Scintillation counters were used to form a trigger which exploited the
charged-neutral-charged signature of $\fullreac$ events by selecting
events in which a 
$\bar{p}$ entered the target vessel, no charged particles left the target 
region and at least one charged particle exited the tracking chambers.  
Figure~\ref{detector_schematic}b  shows the scintillators used to 
require an incident $\bar{p}$ and to veto events in which charged particles 
exit the target area.  Figure \ref{detector_schematic}a shows the two
scintillator hodoscope planes which indicated a charged decay product.

\begin{figure}
\includegraphics[width=8.6 cm,clip]{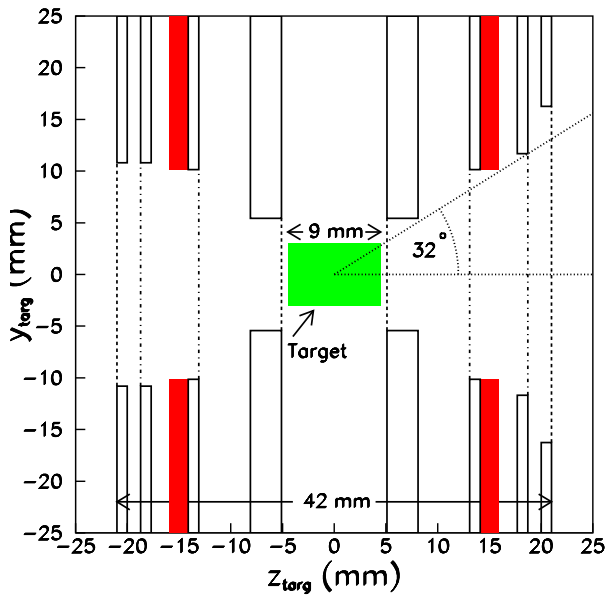}
\caption[]{
\label{Targetregion}
(Color online) Schematic cross-sectional view through the cryostat system (four
concentric vertical cylinders) holding the
frozen spin target (a horizontal cylinder).    Coordinates are
indicated relative to the center of the target.
The antiproton beam entered through the four thin
windows on the left and interacted in the frozen butanol target.
Hyperons escaped with relatively little interaction through the large exit
windows shown on the right.  Dashed lines represent 20 $\mu$m titanium
windows while dash-dot lines represent 40 $\mu$m aluminum windows.
Coils which provide the holding field are shown shaded.
The trigger selected only those events in
which the hyperons survived long enough to escape the cryostat and
pass the veto detectors before decaying.

}
\end{figure}

The most significant change in the apparatus from previous versions of PS185
was the change from a polyethylene active target to a transversely polarized 
frozen spin target\cite{target_thesis}. As described above, it has been shown
\cite{Kent_Brian} that this greatly expands the 
accessible spin information.  The target itself was a solid 9 mm long, 6 mm 
diameter cylinder of frozen Butanol (C$_4$H$_9$OH) doped with 
TEMPO (C$_9$H$_{18}$NO)\cite{target_material}. The 
axis of the cylinder lay along the beam direction.
Figure \ref{Targetregion} shows the nested cryostats and the windows which 
allowed the beam to enter and the $\lbarl$ to exit.  An extremely compact
design was achieved to hold the target at 60 mK within a room-temperature 
outer vacuum vessel of only 2.1 cm  radius.  Such a compact design was 
critical to the success of the experiment as it allowed a significant 
fraction of the $\fullreac$ events of interest to be selected by their neutral
intermediate state.   With a large target vessel the hyperons would have 
decayed internally and the trigger information would have been lost.

The target was polarized roughly every 22 hours by surrounding it by a
superconducting solenoid, with a field of up to 5 T, and pumping 
with microwaves to achieve dynamic
nuclear polarization.  Initial polarizations up to 75.3\% were achieved.  In 
frozen spin mode, with a holding field supplied by an internal solenoid, 
polarization lifetime was roughly 100 hours with beam on target.  The average
polarization over the run was about 62\%.  The polarization direction was
vertical (transverse to the beam) and could be chosen to be positive or 
negative to reduce systematic errors.

 The beam rate on target was approimately $6\times 10^5$ antiprotons
per second.  The integrated beam on target for this measurment was
$1.6\times 10^{11}$ antiprotons.  Scaling by the data acquisition live
time of 79.2\% yields an effective integrated intensity of $1.27\times
10^{11}$ antiprotons.

\section{Data Analysis and Cross Section Results}
\subsection{Event Reconstruction and Kinematic Fitting}

The events of interest, $\fullreac$, are characterized by their two-'Vee' structure
resulting from the decay of the neutral hyperons.  
The first goal of the analysis is to extract 
the small number of candidates for this event topology from the very large
number of events (largely $\bar{n}n$ followed by $\bar{n}$ annihilation)
which satisfy the charged-neutral-charged trigger condition. Less than 0.1\% of
the recorded events were ultimately found to be consistent with the $\fullreac$
hypothesis. The next goal
is to determine which candidate events match the constraints enforced by
energy and momentum conservation on real $\fullreac$ events.  This is done
by kinematic fitting of the event topologies in the tracking chambers,
which are in a region free of magnetic field.
This fit also provides the best estimate of the production
and decay angles of interest and of their correlations.  Additionally,
information from the three drift chambers within the solenoidal magnetic field
is used to determine the sign of the charged particles.  If at least one 
particle's sign can be clearly determined, then the $\Lambda$-$\lbar$ ambiguity
can be resolved and the event can be used in determining spin correlations.

When possible, microstrip information was used in the kinematic fit 
to define the position and 
direction of the incoming $\pbar$.  When this was not possible, average 
beam position and direction were used.  Because of the very small 
emittance of LEAR's adiabatically cooled beam, this was sufficiently precise.

Drift chamber wire positions and time-to-distance calibrations were 
determined empirically from straight track data.  This allowed drift 
chamber hit positions to be accurately determined, typically with 
better than 200 $\mu$m resolution.

For candidate $\fullreac$ events, hit positions were determined from MWPC 
wire positions and from drift chamber wire position and time.  A 
significant walk correction was made for drift chamber hits based upon 
time-above-threshold.  Hits from all planes sharing a common readout 
direction (X,Y,U, or V) were then 
searched for 2-D track projections (X-Z, Y-Z, U-Z, or V-Z).  Drift 
chamber hit positions on a 2-D 
track candidate were iteratively improved by correcting drift-time to
position conversion to reflect the apparent 
slope of the track.  To allow for crossing tracks, a hit could be included 
in more than one 2-D track.  A maximum track angle of 60$^\circ$ was 
allowed in each projection since tracking was inaccurate beyond that 
range.  Losses due to that cut were accounted for as acceptance losses 
as discussed in the Sec.~\ref{fit}.

Combinations of three or more of these 2-D tracks were then considered
as candidate 3-D tracks.  The 
45$^\circ$ rotation of the MWPC projections relative to the drift chamber 
projections helped eliminate spurious combinations.  Confidence-level cuts 
based on $\chi^2$ were used to determine whether sets of 2-D projections 
were consistent with a 3-D track.  Tracks constructed from just two 2-D 
projections were considered only if neither projection could be used to 
form a 3-D track having more projections.

Candidate Vee's were formed from pairs of 3-D tracks having a 
distance-of-closest-approach consistent with zero.  These Vee's were 
rejected if they were not consistent with in-flight decay of a $\Lambda$ 
in the momentum range (471-1161 MeV/c) expected for $\fullreac$.  A single 
3-D track was allowed to be included in more than one Vee to avoid losing a 
real Vee by having a track mis-assigned to  a false Vee.

Pairs of Vee candidates (which did not share any 3-D tracks) were then 
considered as candidates for kinematic fitting.  The pair was first 
tested for rough consistency with the kinematic hypothesis. (e.g. Transverse 
components of momentum,  calculated independently from the topology of
each Vee, should be opposite and equal, within errors.)

For all tracking cuts, simulated events (discussed in Sec.~\ref{mc}) 
were used as a guide in setting confidence-level cuts to avoid cutting 
good events.  The simulation included effects due to finite resolution 
and multiple-scattering.

Kinematic fitting was based upon the fact that the ideal topology of an 
event (neglecting finite resolution, multiple-scattering, and 
interactions) can be completely predicted in terms of 14 parameters:
\begin{list}{$-$\ \ \ }{\itemsep -.05 in}
\item 3 components of beam momentum
\item 3 coordinates of production vertex
\item 2 decay lengths
\item 2 production angles ($\Theta_{c.m.}$ and $\Phi_{c.m.}$)
\item 4 decay angles ($\theta_\Lambda$, $\phi_\Lambda$, 
                $\theta_{\lbar}$, $\phi_{\lbar}$ ).
\end{list}
The production and decay angles are of greatest interest since they 
hold the information on the spin dynamics.  By varying these 14 parameters it 
should be possible to find a hypothesis which is consistent with the 
hits on the observed Vee-pair, within errors.  If satisfactory consistency 
cannot be achieved then the pair can be rejected as not originating 
from $\fullreac$.  If an acceptable fit can be found, then 
it gives the best estimate of the 14 parameters.

When evaluating consistency of a hypothesized set of parameters not 
only the measured track hits were used but also 
a data point was included to represent, with appropriate errors, the
knowledge of the beam energy and the beam direction for that event.
Individual hits (as opposed to tracks) were treated 
as measurements in evaluating the goodness of fit.  The errors on the 
measurements were not treated as being independent, however.
Evaluation of goodness of fit took into account the fact that the errors 
in  hit positions on any given track were correlated because of 
multiple-scattering.  This correlation was increased by the fact that the 
scattering did not always happen uniformly along a track but could be 
greatly increased at points where the particle hit a wire of the tracking 
chambers.  Simulated events were used to study the covariance introduced by 
multiple-scattering and to ensure that the fitting procedure properly 
accounted for it.

A generalization \cite{Kent_thesis} of the Levenberg-Marquardt\cite{Numerical_recipes} method was used to adjust the 
14 parameters to minimize a likelihood statistic (analogous to
$\chi^2$) which accounted for 
covariance due to multiple-scattering. 
The estimated errors on parameters were assigned based on the 
covariance matrix from the fit.
As an example of the accuracy of event-reconstruction, the azimuthal
production angle $\Phi_{c.m.}$was typically determined with an 
r.m.s. error of less
than $0.4^\circ$ (except when it diverged near the poles at 
$\sin(\Theta_{c.m.})=0$).  Also $\cos(\Theta_{c.m.})$ had a mean r.m.s. error of
roughly 0.04 for the worst-case events ($\Theta_{c.m.} \approx 90^\circ$)
falling roughly linearly to less than 0.004 for $0^\circ$ and $180^\circ$ 
scattering.

\begin{figure}
\includegraphics[width=8.6 cm,clip]{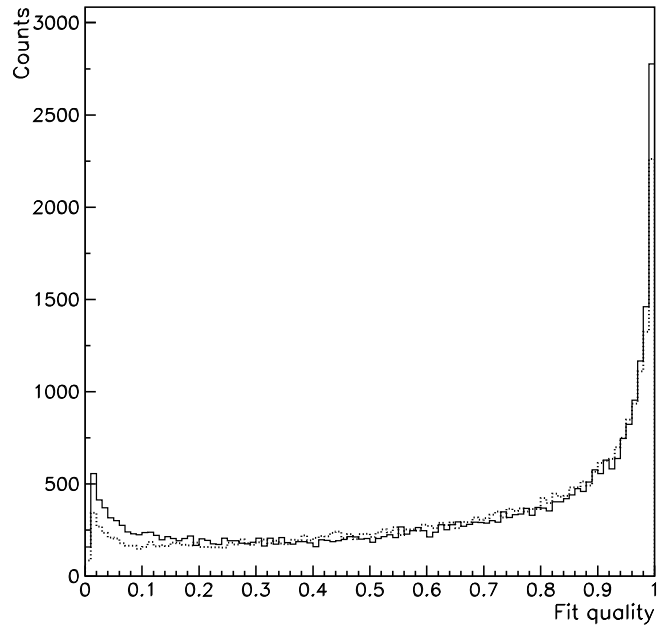}
   \caption{\label{Q_dist}
Distribution of fit-quality, ${\cal Q}$, found by
treating liklihood as if it were distributed with a $\chi^2$
statistic.  Shown dashed is Monte Carlo prediction, normalized to the
data over the interval $.1\le {\cal Q} \le 1$.
 }  
\end{figure}

The fit was heavily over-constrained by the measured topology of the event, 
along with the 12 constraints due to 4-momentum conservation at all three 
vertices.  
 A rough measure of fit-quality, ${\cal Q}$, was calculated by
  treating the likelihood statistic as if it were
  $\chi^2$-distributed and calculating the confidence level.  A flat
  distribution over $0\le {\cal Q}\le 1$ would be expected from a true
  $\chi^2$ statistic.  The actual fit-quality distribution has a large
  peak below ${\cal Q}=.008$ resulting from unrelated 
  background events (such as anti-neutron annihilation) which produce
  large numbers of tracks. This very sharp peak was cut early in 
the analysis and
  will not be included in the discussion of fit-quality which follows.
The remaining {\cal Q} distribution is shown in Figure \ref{Q_dist}.
The peak at high fit-quality ($0.9\le {\cal Q}\le 1.0$) is 
understood as resulting from events in which no
track was substantially deflected by hitting a chamber wire. 
Conversely, the peak a low {\cal Q} results largely from events in
which one or more track suffered significant deflection.
 As shown in Figure \ref{Q_dist},  Monte
Carlo simulation correctly predicted the general shape of the observed
distribution, including the skew towards high {\cal Q} and the 
peaks at large and small ${\cal Q}$.
The small residual differences are believed to result
from imperfect modeling of the multiple scattering and matter
distribution.

  Events with ${\cal Q}<0.1$ were rejected.   This cut not only cleanly
  rejected the spike of background events at very low {\cal Q}, 
 it also rejected a significant fraction of the events from
quasi-free $\lbarl$ production on a bound proton 
in a carbon, oxygen, or nitrogen nucleus in the target.  The efficiency 
of rejection of such events was tested by fitting the $\fullreac$ 
hypothesis to a dataset which was collected using  a pure carbon 
target.  
 These events showed a peak at low ${\cal Q}$, with about
  half the events being eliminated when events with ${\cal
    Q}<0.1$ were cut.  The remaining tail extended 
across the ${\cal Q}$ distribution.
Scaling that data to the number of non-Hydrogen nuclei in the doped 
Butanol target, the quasifree contamination can be estimated 
as 1.7$\pm$0.1\% of the final $\lbarl$ data set.  Since the protons 
involved are unpolarized, this is expected to appear mostly  as a 
dilution of the extracted spin observables.  This effect has been 
included in estimates of the systematic errors.

 The cut was
  expected to reject approximately 9\% of good $\fullreac$ events and
  so would be expected to not significantly bias spin correlations
  unless there was a strong angular variation in the losses.
  Monte Carlo simulated distributions for azimuthal production angle and
  individual decay angles matched those of the actual accepted angular
  distributions, without indication of angular variation of the losses.
  Furthermore, analysis of
  Monte Carlo simulated data with the same analysis cuts 
showed the reconstructed
  spin-correlation and spin scattering matrix parameters to be in
  agreement, within expected statistical errors, with those used to
  generate the Monte Carlo events.

Kinematic fitting in the field-free region cannot distinguish the $\Lambda$ 
from the $\lbar$.  Tracks were extended into the solenoid and the three 
drift chambers there were searched for triplets of hits which could be 
used to assign a charge to at least one of the tracks.  The hits were 
compared to predictions based upon the reconstructed momentum and angle of the 
track and both possible charges.  Expected covariance of the hits, 
which was large because of multiple-scattering in the coils of the 
solenoid, was taken into account.  Any track which was well fit by 
one charge hypothesis acted as a 'vote' for which Vee was the $\Lambda$.  
In principle a single vote was sufficient to resolve the ambiguity and 
properly identify all four tracks.  Events with multiple votes 
occasionally had two tracks which voted differently on which Vee should 
be identified as the $\Lambda$.  
Additionally, 8\% of the events were unusable because they had no 
vote or had conflicting votes.

A na\"ive estimate of the error rate of these individual votes
can be achieved by assuming the
error rate is a constant, uncorrelated with the number of 
votes.  The rate of inconsistency between the votes in two-vote events 
would then imply a 3.1\% error rate on individual tracks.
Assuming that the same error rate applies for those events which 
have only a single vote leads to the estimate that in total 
the $\Lambda-\lbar$ 
identification was interchanged in 1.1\% of the reconstructed events.  

 More careful evaluation however shows that this na\"ive estimate
  under-predicts the actual rate of inconsistency for multiple-vote
  events, indicating a correlation between number of votes and error
  rate.  Without adjustable parameters, the Monte Carlo simulation
  described in the next section makes an excellent prediction of both the
  observed distribution of number of votes per event and the rate of
  inconsistent votes as a function of the number of votes.  This gives
  some confidence in the Monte Carlo prediction that the actual
  fraction of events in which the $\Lambda-\lbar$ identification was
  interchanged is only 0.7\% with no marked dependence on $\Theta_{c.m.}$.
    The smaller value results from the Monte Carlo's prediction of a
  lower error rate for single-vote events than for two-vote events.

Combining these two estimates, a ($0.9\pm 0.2$)\% misidentification
rate was assumed when correcting for contamination as
explained in Sec.~\ref{fit}.
 The estimated error of $\pm 0.2$\% was included in the systematic 
error analysis.  This contamination was negligible in all but the two
most back-angle bins.

A total of 30818 events were successfully kinematically reconstructed 
as $\lbarl$ events with the $\Lambda$-$\lbar$ ambiguity resolved.

\subsection{Monte Carlo Simulation}
\label{mc}

An understanding of the angular dependence of acceptance is critical to 
successful extraction of spin scattering information from angular
 distributions.  The acceptance function  used to extract spin observables
was evaluated using a simulation designed to incorporate 
empirical detector response along with predicted particle interactions.  
The Monte 
Carlo simulation was also used in tuning algorithms and setting cuts to 
optimize tracking, Vee-finding, and fitting.  Simulated events were also used to 
study systematic errors.

The GEANT-based \cite{GEANT} simulation included multiple-scattering, $\delta$-ray 
production, and hadronic interactions.  The latter was especially 
important for the final-state $\pbar$, which has a large annihilation 
cross-section.  The description of the mass distribution included the 
target, cryostat, scintillators, chamber foils, gas, and wires.

The position and response of detectors were determined empirically when 
possible and used as input for simulation.  
Positions of trigger scintillators were determined by tracking 
through the micro-strip detectors using data from dedicated calibration runs 
taken with a thick scatterer upstream of the microstrips.  These data were 
also used to determine tracking chamber positions and to measure their 
efficiencies as a function of track slope and position.  

Simulation of the wire chambers included observed decrease in chamber 
efficiencies near the sense wires, the observed effective size of the 
'dead spot' built into the center of each plane, and the observed 
decrease in efficiency on the neighboring drift chamber sense wire due to 
field distortion at the 'dead spot'.

The effect of the trigger scintillators was an important component of the 
simulation.  Use of trigger scintillators was essential to select out the 
rare events of interest.  The cryogenic target however made it impossible 
to place the scintillators as close to the production point as had been done 
in all previous versions of the PS185 experiment.  Therefore a large 
fraction of all $\fullreac$ events were lost because at least one hyperon 
decay occurred upstream of the veto scintillators.

\begin{figure}
\includegraphics[width=8.6 cm,clip]{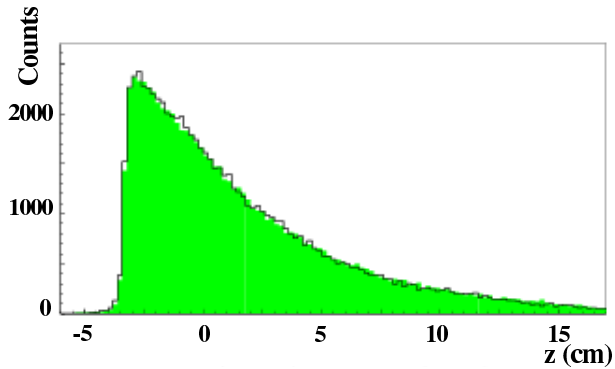}
   \caption{\label{Kent_6.11}
(Color online) Comparison of the hyperon decay vertex $z$-coordinate 
distributions between data (solid line) and simulated data set (shaded).
 }  
\end{figure}

Figure \ref{Kent_6.11} shows a comparison between simulated 
and measured distributions, as an example of a test of the simulation.  
The z component of vertex position of $\Lambda$ decay is shown. 
 The distribution is seen to be well predicted by simulation, including 
the sharp rise due to the position of the veto scintillators.  The
target, as shown in Fig.~\ref{Targetregion}b, is located upstream of
this position but the veto scintillators prevent triggers for events
in which the decay occurs further upstream.

\subsection{Differential Cross Section}

\begin{figure}
\includegraphics[width=8cm,clip]{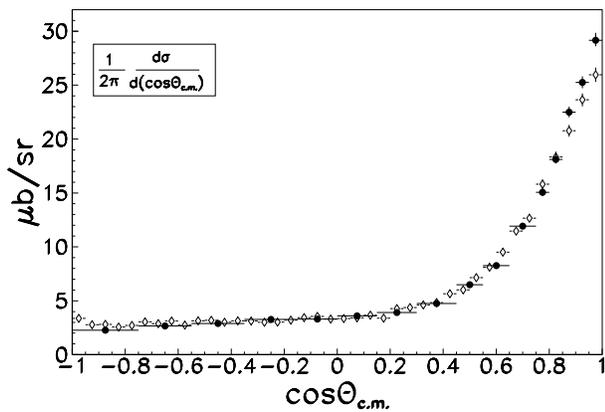}
    \caption{\label{Kent_7.20}
Results for the $\phi$-averaged differential cross section (solid). 
The previous measurement~\protect\cite{Horsti} at $1.642\pGeV$, scaled
by a factor of 1.26 to match present integrated cross section, is 
superimposed (open diamonds). }  
\end{figure}

The $\Phi_{c.m.}$-averaged differential cross section for $\lbarl$ production at 
$p_{\pbar} = 1.637$ GeV/c can be found from the spin-matrix parameters 
\{a,b,c,d,e,g\}, determined as described in the following sections, as

\begin{eqnarray*}
<\bfrac{d\sigma}{d\Omega_\Lambda}>\ &\equiv\  
&\frac{1}{2\pi}\bfrac{d\sigma}{d\cos\Theta_{c.m.}}
=I_0 \\
&= &\onehalf\{ a^2 +b^2 +c^2 +d^2 +e^2 +g^2 \} .
\end{eqnarray*}
The differential cross section so found is shown as dark points in 
Fig.~\ref{Kent_7.20}.  The cross section is essentially identical 
to that found by counting $\lbarl$ events within each $\cos(\Theta_{c.m.})$ 
bin and correcting for mean acceptance over the angular distributions 
found below.

The unequal bin widths seen in Fig.~\ref{Kent_7.20} were chosen to 
give roughly equal statistics of observed $\lbarl$ events in each of sixteen
$\cos(\Theta_{c.m.})$ bins.  This allowed a stable fit of the eleven 
spin-matrix parameters to be performed in each bin.  Uniform binning 
would have sacrificed forward-angle $\cos(\Theta_{c.m.})$ resolution or 
resulted in low-statistics fits at back angles leading to instability 
and large errors.  The cross section is discussed here, in advance of the 
explanation of the fitting procedure, to motivate this choice 
of $\cos(\Theta_{c.m.})$ bins, which applies to all subsequent discussion 
of the matrix parameters.

The open points in Fig.~\ref{Kent_7.20} show a renormalized version 
of a previously published result \cite{Horsti} from an earlier version 
of the PS185 experiment using unpolarized polyethylene target cells 
surrounded by scintillator.  That measurements were made at $p_{\pbar}=1.642$ 
GeV/c which is very nearly the same beam momentum as the present 
data, $p_{\pbar}=1.637$ GeV/c.  For the purpose of comparison of shapes, 
the older result has been scaled up by a factor of 1.26 to match the 
integrated cross section of $81.1 \pm 0.5 ^{+5.8}_{-7.5}\,\,\mu b$ measured 
in the present experiment (where the first error is statistical 
and the second is systematic).  The systematic errors assigned to the 
present measurement are larger than on earlier ones because the 
cryogenic polarized target introduced larger uncertainties in target 
thickness, target position relative to trigger counters, and hadronic 
interactions.  As described in Sec.~\ref{syserr}, these systematic 
errors have been realistically estimated so an explanation is required 
for the apparent discrepancy between the present and previous 
determinations of total cross section.  The normalization discrepancy 
in early PS185 results has already been described in an earlier 
publication \cite{threshold}.  Hadronic interactions and multiple
scattering were not 
included in the custom-written Monte Carlo simulation code used to 
determine acceptance corrections in early analyses of PS185 data 
\cite{old_PS185,Horsti}.  Versions of a GEANT-based simulation 
have been used in calculating acceptances for more recent results 
\cite{threshold,recent_PS185}.  Inclusion of hadronic 
interactions and multiple scattering increased the estimated 
yield by 8-12\%.  With a 
$10\pm2$\% adjustment in normalization, the older result gives 
$\sigma= 70.4\pm 0.4 \pm 2.2\,\mu b$ which roughly agrees with the present 
result within errors.
The spin-correlations are
insensitive to any systematic error on overall normalization.

Comparison of the two data sets in Fig.~\ref{Kent_7.20} shows some 
difference in shape of the distribution at forward angle.  The present 
analysis took into account correlated errors due to track deflections 
caused by multiple-scattering.  This allowed events to be reconstructed 
which might otherwise have been lost.  The older analysis did not 
allow for multiple-scattering in event reconstruction.  The lost events 
would not have been compensated by acceptance corrections since 
multiple-scattering was not included in the Monte Carlo simulation.  
Furthermore, estimates of expected losses, found by disabling the 
multiple-scattering correlation in fitting in the present analysis, 
indicate that the effect 
is greatest at forward and back angles, where hyperon momenta are low.  
The sharper peak at forward angles seen in the present data is 
therefore believed to be accurate while the older data is slightly 
distorted by multiple-scattering losses.

\subsection{Fitting Spin-scattering Matrix Parameters}
\label{fit}

For each bin in $\Theta_{c.m.}$, Eq.~(\ref{eqn_b}) represents the 
distribution of events across a 5-dimensional space of angles ($\Phi_{c.m.},\ 
\cos(\theta_{\lbar}),\ \phi_{\lbar},\ \cos(\theta_\Lambda),\ \phi_\Lambda$)
which will be represented by a 5-dimensional vector, $\vec{v}$ for
notational convenience.  With only roughly 2000 events in each $\Theta_{c.m.}$
bin, performing a simple $\chi^2$-fit by subdividing each of the five 
coordinates into bins is excluded.  An unbinned maximum-likelihood fitting 
technique was  employed.  This is often called simply 'maximum-likelihood 
fitting' a name which fails to
distinguish from other fit methods such as $\chi^2$ minimization,
which also correspond to a 
maximum likelihood.

Unbinned fitting is a limiting case of fitting with
Poisson statistics.  If the data were binned into $K$ bins with $n_k$
being the number of events in bin $k$, then the likelihood of the
observed data set would be
\[{\calL} = \prod_{k=1}^{K}e^{-\phi(\vec{v}_k,\vec{a})\Delta\vec{v}_k}
\{ \frac{(\phi(\vec{v}_k,\vec{a}) \Delta\vec{v}_k)^{n_k}}{n_k!}\}\]
where $\phi(\vec{v}_k,\vec{a})$ is the probability density function which 
depends on the eleven real parameters of the spin-scattering matrix, here 
represented as the 11-dimensional vector $\vec{a}$.  The volume of the k'th 
5-dimensional bin is $\Delta\vec{v}_k$.
If the number of bins, $K$, now becomes large compared to the number
of events, $N$, then $n_k$ will be zero for most bins and will be unity for
$N$ bins.  In this limit the likelihood becomes
\[{\calL} = \prod_{k=1}^{K}e^{-\phi(\vec{v}_k,\vec{a})\Delta\vec{v}_k}
\prod_{j=1}^{N}\phi(\vec{v}_j,\vec{a}) \Delta\vec{v}_j\]
where the second product runs over only the occupied bins.
  Neglecting background contamination, $\phi$ is the intensity
  distribution, described by Eq.~(\ref{eqn_b}), scaled by the 
integrated luminosity of the experiment, 
the solid angle corresponding to the $\Theta_{c.m.}$-bin considered, and the
average acceptance probability for coordinates $\vec{v}_k$, $\vec{v}_k$.

Taking the log of ${\calL}$ introduces a sum of $\ln(\Delta\vec{v}_j)$ terms
which would diverge in the limit of infinitesimal bin size.  Since these terms 
are independent of $\vec{a}$, discarding them 
does not change the value of $\vec{a}$ which maximizes the likelihood.
Let ${\cal M}^\prime$ represent $-2\ln({\cal L})$ with these terms discarded.
Then in the limit of infinitesimal bin size,
\[{\cal M}^\prime = 2\int \phi(\vec{v},\vec{a})\ d\vec{v} - 
2\sum_{j=1}^{N}\ln(\phi(\vec{v}_j,\vec{a})) \]

Maximization of the likelihood for a data set \{$\vec{v}_1$ ... $\vec{v}_N$\} 
can be accomplished by finding the parameters, $\vec{a}$ which minimize 
${\cal M}^\prime$.  Writing $\phi(\vec{v},\vec{a})$ as $\mu(\vec{v},\vec{a})
A(\vec{v})$ where A is the acceptance and $\mu$ is the remainder of the 
probability density function (which, neglecting background
contamination, would be 
$I_{final}(\Theta_{c.m.},\Phi_{c.m.},\hat{k}^{\bar{p}},\hat{k}^p)$ scaled by 
the integrated luminosity and the solid angle),
\begin{eqnarray*}
{\cal M}^\prime = 2\int \mu(\vec{v},\vec{a})A(\vec{v})\ d\vec{v} &- 
&2\sum_{j=1}^{N}\ln(\mu(\vec{v}_j,\vec{a})) \\
&-&2\sum_{j=1}^{N}\ln(A(\vec{v}_j)) 
\end{eqnarray*}

The last term, which is independent of $\vec{a}$, can be 
discarded without affecting the position of the 
minimum.  This gives the final function to be minimized, which will 
be called ${\cal M}$.

The first term of ${\cal M}$, which incorporates the acceptance
function, could in principle be 
evaluated by Monte-Carlo integration (with each simulated event being processed
by the analysis routines to see whether it would be successfully 
reconstructed).  However, minimization of ${\cal M}$ would then require
prohibitive re-evaluation of this term for each new set of parameters 
$\vec{a}$ being tested.  Fortunately the structure of $I_{final}$
in Eq.~(\ref{eqn_b}) permits a great simplification.  Each of the twenty 
terms of Eq.~(\ref{eqn_b}) can be written as a product with an 
$\vec{a}$-dependent dynamic term (containing the Q's and $I_0$) multiplying 
a purely geometric term, $G_i(\vec{v})$, which depends on $\vec{v}$
 but not on the parameters $\vec{a}$.  So $\mu$ can be written as 
\[\mu(\vec{v},\vec{a}) = \sum_{i=1}^{20} D_i(\vec{a})G_i(\vec{v})\]
allowing the first term of ${\cal M}$ to be simplified since 
\begin{eqnarray*}
\int \mu(\vec{v},\vec{a})A(\vec{v})\ d\vec{v} &= 
&\sum_{i=1}^{20} D_i(\vec{a})\int G_i(\vec{v}) A(\vec{v})\ d\vec{v}
\\
&=&\sum_{i=1}^{20} D_i(\vec{a}) W_i
\end{eqnarray*}
where the weights $W_i$ are the moments of the acceptance which 
are independent of $\vec{a}$ and 
so need to be evaluated only once by Monte-Carlo integration.  To
perform this integration the simulation was used to 
generate events uniformly distributed 
in $\vec{v}$-space without weighting to match the observed spin-correlations.
The fitting procedure is then to search for $\vec{a}$ which minimizes
\[{\cal M} = 2   \sum_{i=1}^{20} W_i D_i(\vec{a}) -
2\sum_{j=1}^{N}\ln(\mu(\vec{v}_j,\vec{a})) \]

As explained above, an estimated $(0.9\pm 0.2)$\%
rate of mis-identification of $\Lambda$ and $\lbar$
causes contamination of $\Theta_{c.m.}$ bins with data from the supplementary 
angle.  Because the differential cross section is strongly forward-peaked, 
this is a negligible effect except for the two most back-angle bins.
 The predicted 
mis-identification rate at forward angles leads to an 8\%
  contamination of the 
most back-angle bin.  To allow for this, the $\mu(\vec{v})$ used 
in back angle bins was not simply $I_{final}(\vec{v},\vec{a})$ scaled by
the integrated luminosity and solid angle.  Rather, 
$I_{final}(\vec{v},\vec{a})$
was replaced by an appropriately weighted linear combination of 
$I_{final}(\vec{v},\vec{a})$ and the background term 
$I_{final}(\vec{v}_{reversed},\vec{a}_{forward})$ where $\vec{v}_{reversed}$
is the corresponding point in $\vec{v}$-space reached by reversing $\Lambda$
and $\lbar$ identification and $\vec{a}_{forward}$ is the set of parameters 
which applies for the forward-angle bin.  Fitting of forward-angle bins
was carried out first to allow this contamination to be correctly modeled when
fitting the back-angle bins.

Minimization in the 11-dimensional $\vec{a}$-space was accomplished by the
Polak-Ribi\`ere conjugate gradient method \cite{Numerical_recipes}.  To guard 
against false minima, each minimization was carried out multiple times 
with different randomly-chosen starting points.  For most 
$\cos(\Theta_{c.m.})$ bins a common minimum was found in every search.  
In the worst case the fit converged on false minimum less than 65\% of 
the time.  Additionally, Monte Carlo 
simulated data sets were generated with similar statistics and 
spin-correlations to the actual data.  Fitting these simulated data sets 
demonstrated the robustness of the technique for converging on the
proper minimum and allowed determination
of a scale of confidence level for numerical values of ${\cal M}$.  Unlike
$\chi^2$, ${\cal M}$ has no {\it a priori} expected value for good fits
because it is arbitrarily offset from the true log-likelihood.  There was
no indication that ${\cal M}$ values obtained for the real data were 
systematically higher than what was expected for a good fit, based on
values obtained for corresponding simulated data sets.

Use of the curvature matrix, based on the assumption of parabolic behavior of
${\cal M}$ near the minimum, was found to give inaccurate estimates
of the errors on the $\vec{a}$ parameters.  Errors were instead determined by 
using a 'brute-force' search of the space around the minimum to find the 
maximum possible change in each parameter (in conjunction with changes in 
all other parameters) consistent with an increase of less than one in 
the value of ${\cal M}$.  Since ${\cal M}$ differs from $-2 \ln({\cal L})$
only by a constant offset, a given change in ${\cal M}$ has the same 
interpretation as the equivalent change in $-2 \ln({\cal L})$.  These 
maximum-acceptable (positive- and negative-) changes in each parameter will be
referred to as the '$1\sigma$' errors since they cover the confidence interval
which would be covered by $1\sigma$ in the case of Gaussian errors.  Similarly,
a '$2\sigma$' error bound on each parameter was found by finding the maximum
possible (positive or negative) change in each parameter for which 
${\cal M}$ would exceed its minimum value by less than $4$.

Once the best-fit parameters $\vec{a}$ have been found, $I_0$ and all the Q's 
can be determined (even those Q's not directly measurable in an experiment 
with unpolarized beam and a transverse target polarization).  But, since the 
errors on the parameters are highly correlated, the error on each 
spin-correlation cannot be determined by simple lowest-order error propagation.  
Even if the correlation matrix were determined, error propagation would 
be unreliable.  A far superior estimate of the error on each 
spin-correlation comes from the same 'brute-force' search for 
$\Delta {\cal M} = 1$ and $\Delta {\cal M} = 4$ regions. 
 At each point, all quantities of interest, such as spin-correlations, 
were calculated.  The maximum positive and negative excursions of each 
such quantity from its best-fit value, consistent with  
$\Delta {\cal M} < 1$ and  $\Delta {\cal M} < 4$, were taken 
to be the '$1\sigma$' and '$2\sigma$' errors, respectively, on that calculated
quantity.

\subsection{Systematic Errors}
\label{syserr}

Overall normalization errors would not affect measured spin-correlations 
but would directly change $I_0$ and equivalently the $\Phi_{c.m.}$-averaged 
differential cross section presented in Fig.~\ref{Kent_7.20}.  Since the 
focus of the experiment was on spin-correlations, overall normalization 
was not controlled as carefully as other aspects of the experiment.  The 
length of the cylindrical frozen Butanol target was measured to be 
$9.0\pm0.5$ mm.  An upper limit on the misalignment of the target 
axis relative to the beam was $20^\circ$ which could increase effective 
target thickness by up to 6\%.  The uncertainty in effective position of 
veto scintillators relative to the target was estimated at 1 mm which 
was found, through Monte Carlo simulation, to cause less than 4\% 
uncertainty in normalization.  Statistical error in the estimate of 
quasi-free contamination introduced a negligible systematic error.  
The overall fractional systematic uncertainty, estimated by adding these 
contributions in quadrature, is ${+7.1\%}/{-9.3\%}$.  This gives the
systematic error, quoted above, on the total cross section, 
$\sigma = 81.1 \pm 0.5 ^{+5.8}_{-7.5} \mu b$.  This same fractional
systematic error applies to the bins of $<\frac{d\sigma}{d\Omega}>$ in
Fig.~\ref{Kent_7.20}.  While the normalization error cancels in
spin-observables, all spin-matrix parameters would scale by the
square-root of any overall normalization factor.  Since this is a
common factor on all terms, it is not included in the error band
assigned to each of these parameters.  It should be remembered that an
overall normalization error of  ${+3.5\%}/{-4.8\%}$ applies to the
spin-matrix parameters.

The reconstructed angles for each event, $\vec{v}_k = (\Phi_{c.m.},\ 
\cos(\theta_{\lbar}),\ \phi_{\lbar},\ \cos(\theta_\Lambda),\
\phi_\Lambda$),  are not known exactly but are extracted with known
errors and correlations.  The method of unbinned maximum-likelihood
fitting treats each event as a precise point in the 5-dimensional
$\vec{v}$-space and does not incorporate a method of allowing for the
finite errors on these points.  Neglecting these finite errors
introduces a source of systematic error in addition to the statistical
error discussed above.  The method of estimating these systematic
errors was based on the use of simulated data with errors and event
statistics similar to the actual data.  These simulated data sets
were generated based on spin-correlations chosen to nearly match those
found in the data.  Simulated events were kinematically fit giving
$\vec{v}_k$'s with errors similar to the real data.  Unbinned
maximum-likelihood fits were then used to extract best-fit values of
the spin-transfer matrix parameters, which were then used to calculate
spin-correlations and all other variables of interest, as described in
the next section.  The fitting
process was then repeated using the ideal $\vec{v}_k$'s which had
been used by the Monte Carlo to generate the simulated data.  The
differences between best-fit values of each variable fit to the
simulated data and the best-fit values fit to the ideal $\vec{v}_k$'s was
a measure of the systematic error in that variable because of the
finite resolution of the $\vec{v}_k$'s.  Changes in each variable of
interest were calculated independently, rather than using
error-propagation to extract the expected effect.  Ideal fits were
compared to fits using kinematically fit finite-resolution data for
ten different simulated data sets.  Additionally, since kinematic
fitting was quite time-consuming, the statistics of this study were
augmented by generating an additional 20 data sets by simply smearing
each ideal $\vec{v}_k$ by its estimated resolution rather than simulating an
event and operating the full analysis chain on it.  Estimates of
systematic error due to finite resolution were thus found as the
r.m.s.~shift from the zero-resolution value for each variable of
interest in each $\Theta_{c.m.}$ bin.  This contribution typically
dominated the overall systematic error estimates which are shown with
each variable in the next section. 

The non-hydrogen nuclei in the target were unpolarized
so quasi-free events mis-identified as free $\fullreac$
events will exhibit no correlation to target polarization direction.
The effect of the small quasi-free contamination can therefore be
estimated by introducing an appropriate fraction of
isotropically-distributed simulated events in place of some of the
events of a simulated data set.  The estimated quasi-free
contamination (from analysis of carbon-target data) was typically only
about 1\% but rose to 3\% at farthest back-angle.  Systematic errors
for each variable of interest were again found by determining the
r.m.s.~shift of that variable caused by the simulated contamination.
This small contamination generally caused smaller systematic errors
than did the effects of finite resolution.

Several precautions were taken to reduce systematic effects due to
target polarization.  The direction of target polarization was
reversed during data collection to cancel systematic effects due to
any up-down asymmetry of the detectors.  Average target polarization
was measured by Nuclear Magnetic Resonance measurements at the 
beginning and end of each data-collection
period (between target re-polarizations) to improve estimates of its
value at intermediate times.  Data-collection periods were typically
limited to a quarter of the polarization lifetime to keep the
polarization high and to improve accuracy of estimated polarization.
The probability density function used to weight each measurement in
the unbinned fit was not based upon an average polarization but on the
best estimate of the polarization at the time that event was recorded.

An additional systematic error results from
uncertainty in initial polarization and from possible inhomogeneities
in target relaxation.  The fractional error in measured target
polarization is estimated at ${\Delta P}/{P} = 2.3\%$.  The
average polarization measured at the end of a data-collection period
could not determine whether depolarization was non-uniform due to beam
heating.  However the target was kept cold enough that relaxation time
was not strongly temperature-dependent.  Inhomogeneity of
depolarization is estimated to contribute an uncertainty of at most 
${\Delta P}/{P} = 3.8\%$ just before repolarization.  These errors
combined to give a worst-case estimate of ${\Delta P}/{P} = 4.5\%$
at the end of the data-collection period. 

The maximum systematic error due to polarization was estimated by
shifting polarizations by 4.5\% and determining the size of the shift
of extracted observables of interest.  This was found to be a smaller
error than that due to resolution or quasifree background.  All three
effects were added in quadrature bin-by-bin for all variables of
interest to obtain the overall systematic error estimate which is
plotted as an error band at the bottom of plots of each variable of interest.

\section{Spin Correlation Results}

\begin{figure*}
\includegraphics[width=17.5 cm,clip]{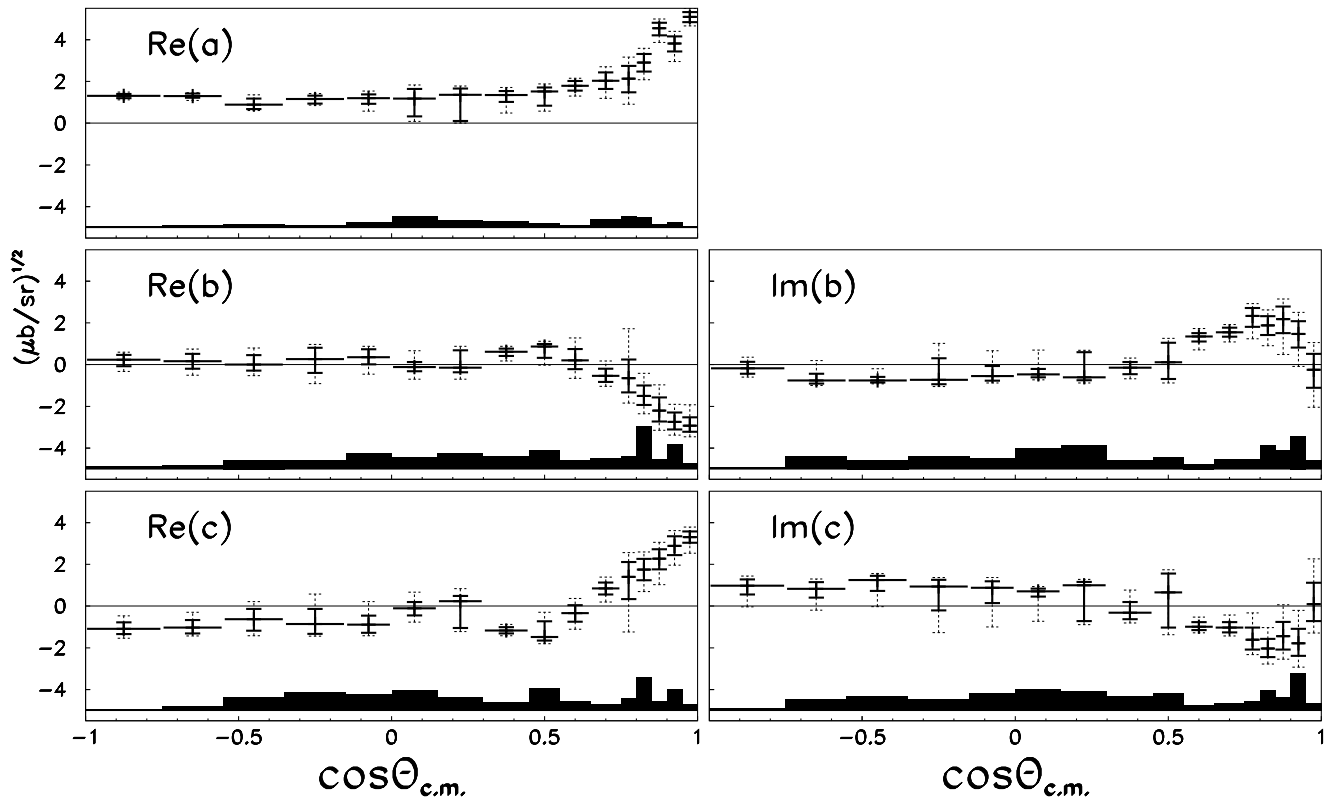}
    \caption{\label{result_1}
Fit results for spin matrix parameters a, b, and c.  The arbitrary
phase is chosen by constraining the parameter $a$ to be real and non-negative.
Statistical errors are shown on each data point, with $2\sigma$ error bars
superimposed (dashed).  The estimated systematic error
width is shown at the bottom of each plot (dark-shaded region). }  
\end{figure*}

\begin{figure*}
\includegraphics[width=17.5 cm,clip]{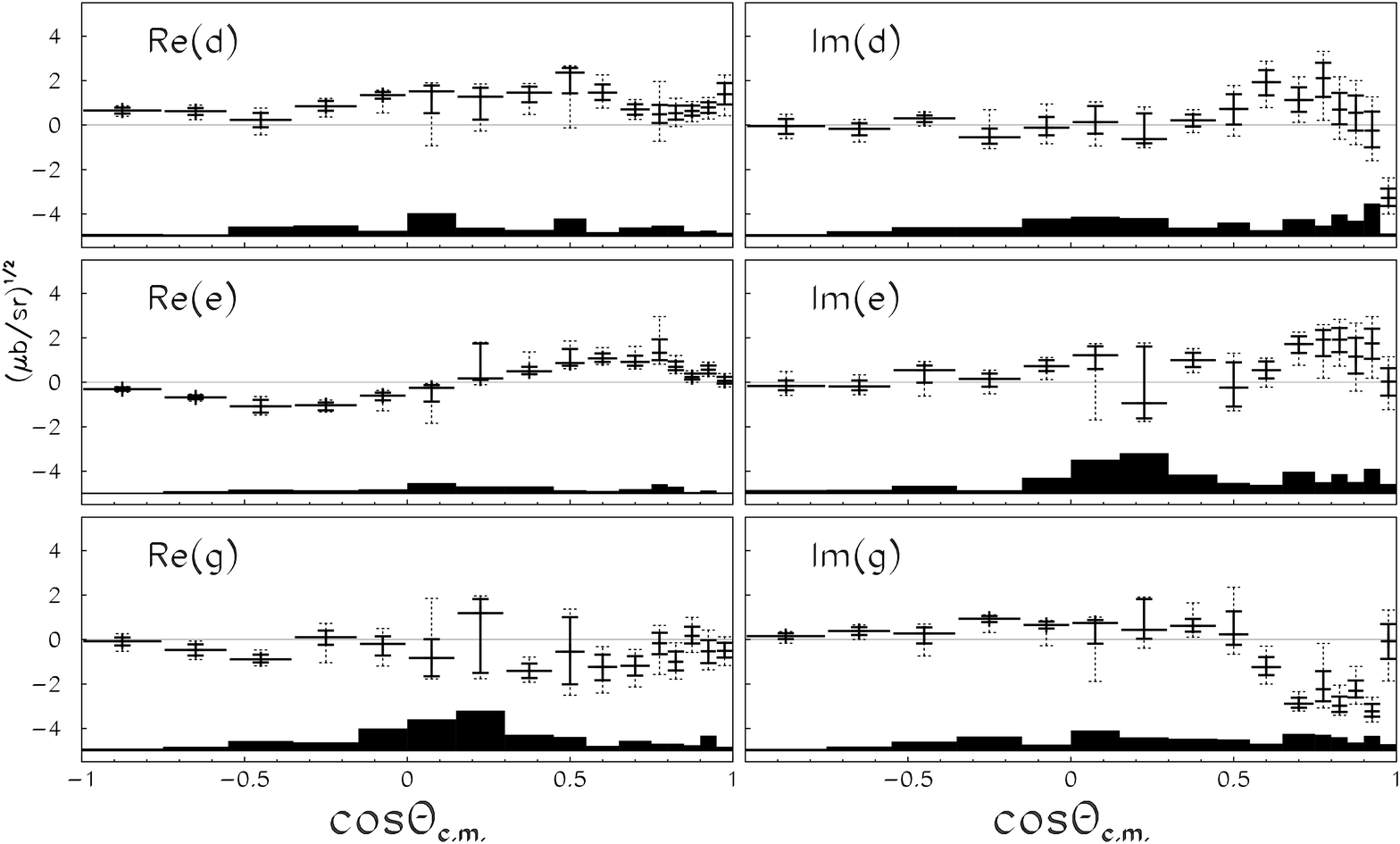}
    \caption{\label{result_2}
Fit results for spin matrix parameters d, e, and g.
Statistical errors are shown on each data point, with $2\sigma$ error bars
superimposed (dashed).  The estimated systematic error
width is shown at the bottom of each plot (dark-shaded region). }  
\end{figure*}

Figures \ref{result_1} and \ref{result_2} show the best-fit values of the 11 parameters
of the spin-scattering matrix  fit to the data in bins of $\Theta_{c.m.}$.
Parameter 'a' is chosen to be real (and non-negative) so Im(a) is zero.  
On each point the dark bar indicates the 
'$1\sigma$' error range while the light bar represents the '$2\sigma$' error.
Because of the actual shape of the ${\cal M}$-hypersurface, some of the 
error bars are highly asymmetric and the '$2\sigma$' error is often very
different from twice the '$1\sigma$' error.  The black band at the
bottom of each plot indicates the 
estimated systematic error.  Overall normalization error is not included in
the systematic-error band.   These,
and other results presented below are available in table form\cite{EPAPS}.

\begin{figure}
\includegraphics[width=8.6 cm,clip]{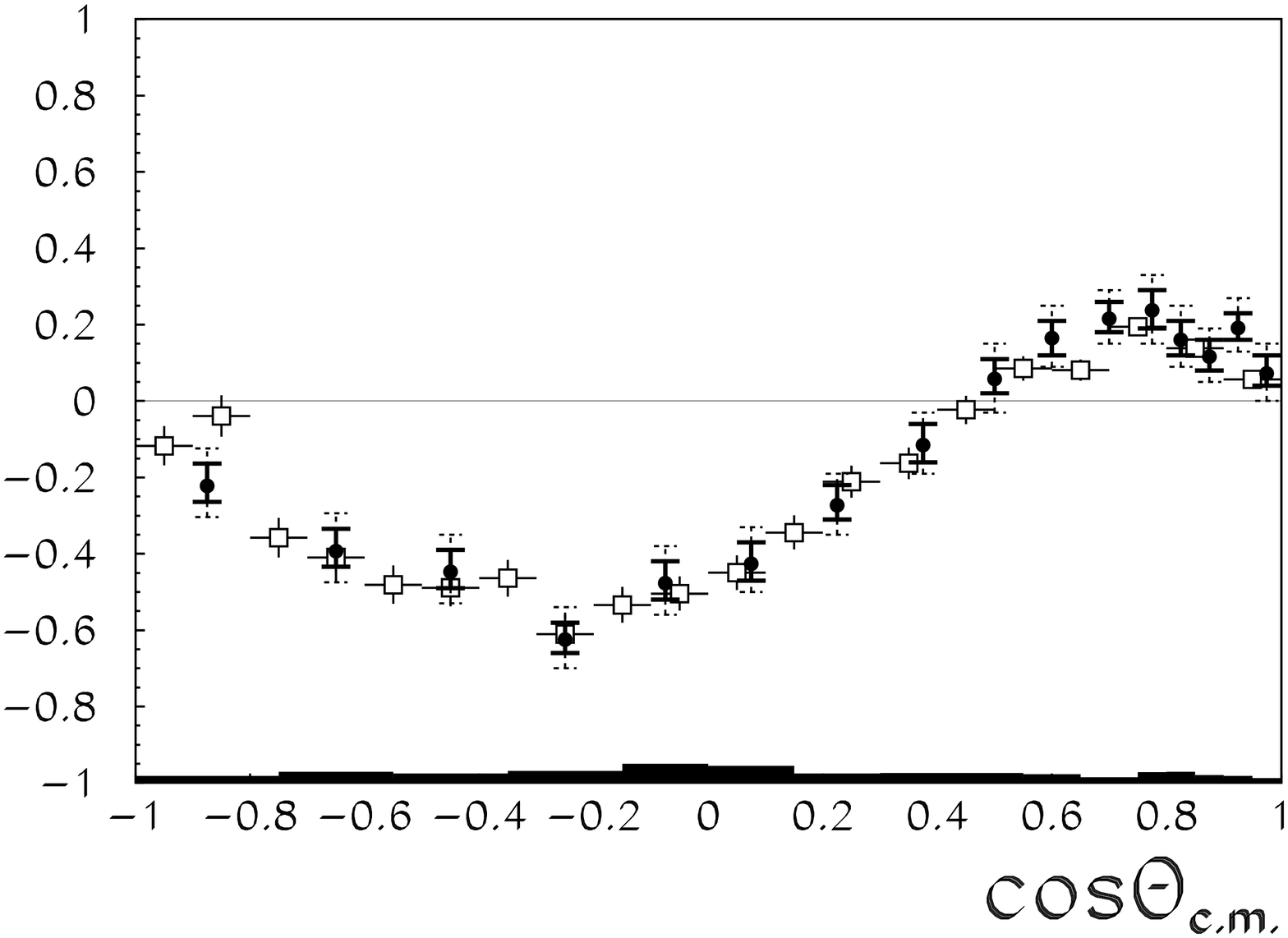}
    \caption{\label{Kent_7.21}
Current results for polarization, $Q[n_\Lambda] = Q[n_{\lbar}]$ (filled circles).
The previous measurement~\protect\cite{Horsti} at $1.642\pGeV$ is
superimposed (open squares).
Statistical errors are shown on each current data point, with $2\sigma$ 
error bars superimposed (dashed).  The estimated systematic error
width is shown at the bottom of each plot (dark-shaded region). }  
\end{figure}

\begin{figure*}
\includegraphics[width=17.6 cm,clip]{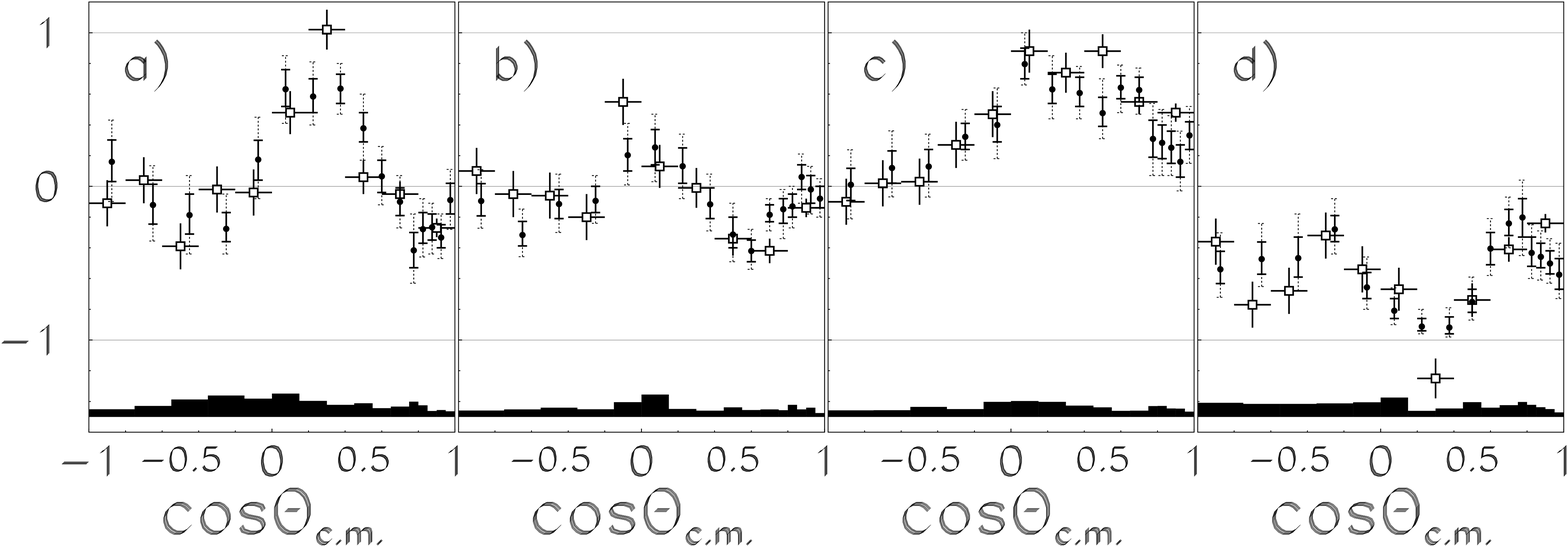}
    \caption{\label{Kent_7.22}
Current results for previously measured spin correlations
between $\lbar$ and $\Lambda$ (filled circles).
Previous measurements~\protect\cite{Horsti} are superimposed (open squares).
Statistical errors are shown on each current data point, with $2\sigma$ 
error bars superimposed (dashed).  The estimated systematic error
width is shown at the bottom of each plot (dark-shaded region). The
spin correlations shown are 
a) $Q[m_{\bar{\Lambda}},m_\Lambda]$,
b) $Q[m_{\bar{\Lambda}},l_\Lambda]$,
c) $Q[n_{\bar{\Lambda}},n_\Lambda]$,
and d) $Q[l_{\bar{\Lambda}},l_\Lambda]$.
}  
\end{figure*}

Table \ref{tab_spin_corr} shows how each of the 19 directly measurable 
spin-correlations (and $I_0$) can be calculated from the parameters
 $\vec{a}$ for the spin-scattering matrix.  While the spin-correlations 
can be calculated directly from the best-fit values of the parameters, as 
explained above, their errors cannot be found by propagation of the errors 
on the parameters.  Results for $I_0$ (= $<\frac{d\sigma}{d\Omega}>$ ) have 
already been shown in Fig.~\ref{Kent_7.20} above.
Results for the polarization $Q[n_\Lambda] = Q[n_{\lbar}]$, 
often denoted $P_\Lambda$
and $P_{\lbar}$, are shown in Fig.~\ref{Kent_7.21}
as filled circles with '$1\sigma$' and '$2\sigma$' errors indicated.  Also
shown are results from the previous PS185 measurement \cite{Horsti} at 
$p_{\pbar} = 1.642$ GeV/c which can be seen to be in good agreement.
Similarly Fig.~\ref{Kent_7.22} shows those 2-spin correlations which can be 
measured without a polarized target.  These correlations of the spins
of the final-state $\Lambda$ and $\lbar$ are commonly denoted $C_{mm}$,
$C_{ml}$, $C_{nn}$,
and $C_{ll}$.  Again good agreement is seen with the 
previous results.

\begin{figure}
\includegraphics[width=8.6 cm,clip]{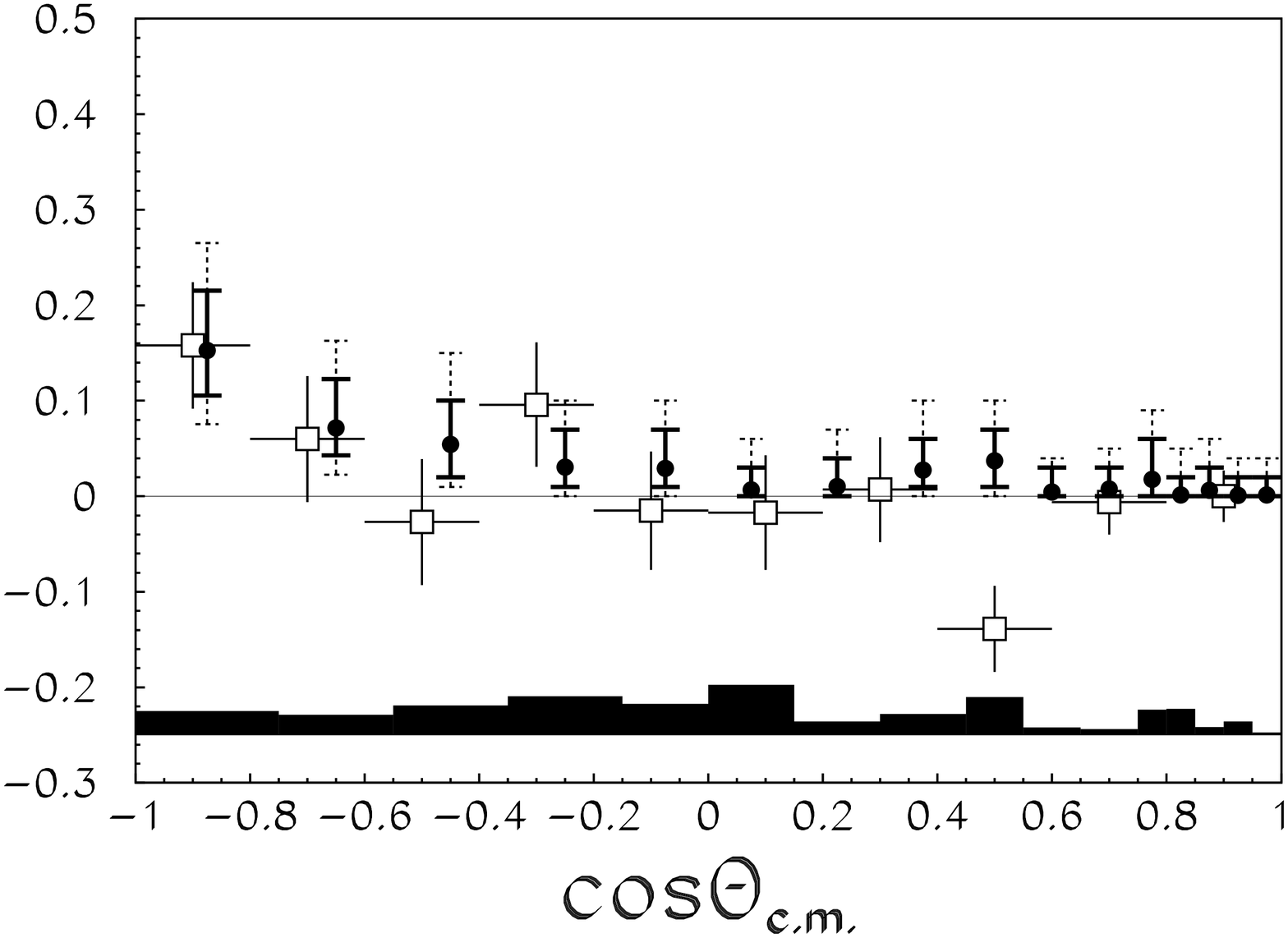}
    \caption{\label{Kent_7.23}
Current results for singlet fraction
$\text{S}_{F}$ (filled circles).
The previous measurement~\protect\cite{Horsti} is superimposed (open squares).
Statistical errors are shown on each current data point, with $2\sigma$ 
error bars superimposed (dashed).  The estimated systematic error
width is shown at the bottom of each plot (dark-shaded region). }  
\end{figure}

A particular combination of these observables, which has direct physical 
interpretation, is the singlet fraction (the fraction of $\lbarl$ pairs which
are produced in a spin-singlet state) which can be written as
\begin{equation}
\label{sf_1}
S_F = \frac{1}{4} (1 - Q[n_\Lambda,n_{\lbar}] + Q[m_\Lambda,m_{\lbar}] + 
Q[l_\Lambda,l_{\lbar}] )
\end{equation}
This can be calculated directly from the spin-scattering matrix parameters as
\begin{equation}
\label{sf_2}
S_F = \frac{|b-c|^2}{4 I_0} 
\end{equation}
This is shown in Fig.~\ref{Kent_7.23}.  Again the errors have been assigned 
by directly determining the limits of change in $S_F$ for a maximum 
acceptable change in log-likelihood.  Previous PS185 results 
\cite{Horsti} at $p_{\pbar} = 1.642$ GeV/c are shown for comparison.  The 
earlier results were determined from Eq.~(\ref{sf_1}) using 
spin-correlations which 
had been separately extracted from the data.  Unphysical negative values 
could then occur as a result of statistical fluctuations or heightened
sensitivity to systematic errors in the linear combination of
observables.  The present results were extracted using 
Eq.~(\ref{sf_2}) and so are constrained to be non-negative throughout the
range of  
their error bars.  It is interesting to note that the often-accepted empirical
rule that $S_F = 0$ for this reaction is clearly broken at back angles.

\begin{figure*}
\includegraphics[width=17.6 cm,clip]{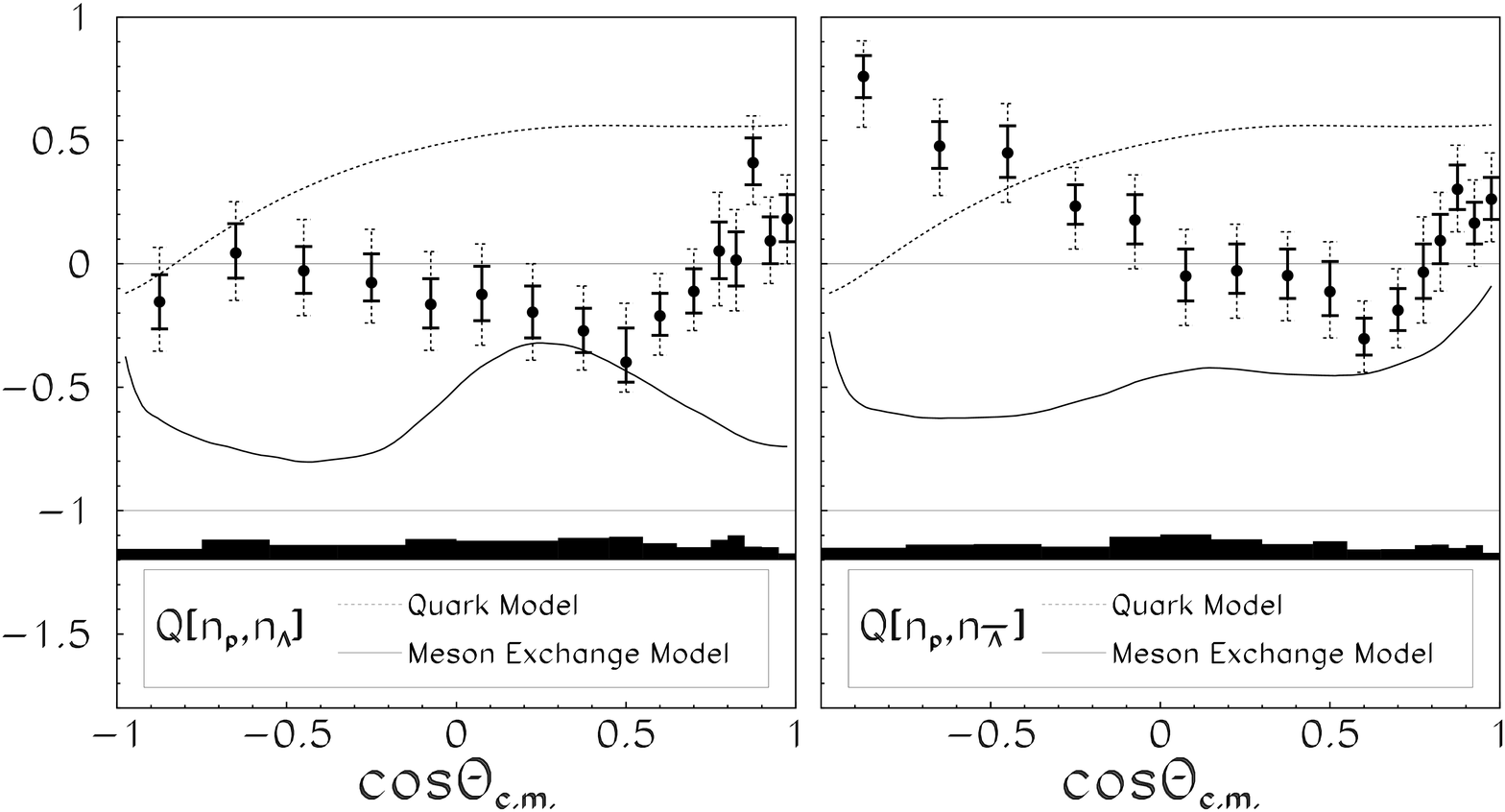}
    \caption{\label{Kent_8.2}
Results for spin transfer observables 
$Q[n_p,n_\Lambda]$ (often called depolarization, $D_{nn}$)
and $Q[n_p,n_{\lbar}]$ (often called spin-transfer, $K_{nn}$)
at $1.637\pGeV$, compared to MEX model 
prediction~\protect\cite{MEX_pred} 
(solid) and QG model prediction~\protect\cite{QG_pred}
(dashed) at $1.642\pGeV$.  Statistical error bars are shown on each
data point, with $2\sigma$ error bars superimposed (dashed).
The estimated systematic error width of the measurement is
shown at the bottom of each plot (dark shaded region). }  
\end{figure*}

 Figure \ref{Kent_8.2} shows results for $Q[n_p,n_\Lambda]$
and $Q[n_p,n_{\lbar}]$ conventionally denoted as $D_{nn}$ and $K_{nn}$, 
respectively.  These results, which have already been published in 
\cite{Bernd_et_al}, are seen to disagree strongly with predictions from 
both meson-exchange \cite{MEX_pred} and quark-gluon \cite{QG_pred} models.
While these results have been published, the present paper is the first
to document the details of the technique used to extract them.
Measurement of these spin-correlations was the main goal of this experiment
because two competing classes of models made differing firm predictions
for these observables.  While both classes of model had enjoyed success in
explaining the observations made with unpolarized targets, these results 
suggest that additional dynamics will have to be included into the models.
The wealth of additional spin-dynamics information presented below may 
help constrain and test refinements made to match the surprising
results in these two spin-correlations.

\begin{figure*}
\includegraphics[width=17.6 cm,clip]{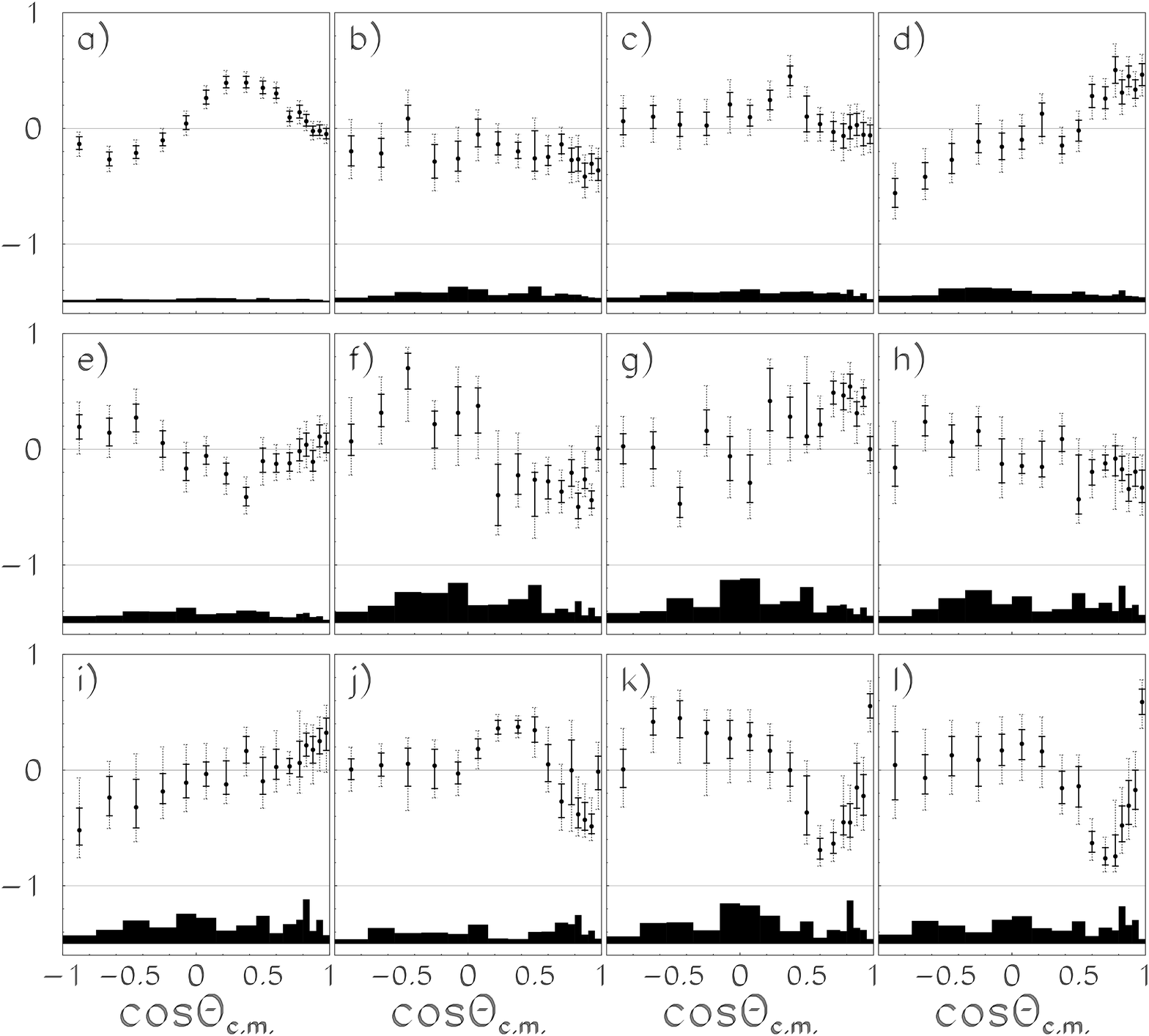}
    \caption{\label{others}
Results for the twelve additional spin observables which 
appear directly in the measured angular distribution. 
Statistical and systematic error estimates are displayed, as above. 
The spin observables displayed are a) $Q[n_p]$,
  b) $Q[m_p,m_\Lambda]$, 
 c) $Q[m_p,l_\Lambda]$, 
d) $Q[m_p,m_{\bar{\Lambda}}]$, 
e) $Q[m_p,l_{\bar{\Lambda}}]$, 
f) $Q[m_p,m_{\bar{\Lambda}},n_\Lambda]$, 
g) $Q[m_p,n_{\bar{\Lambda}},m_\Lambda]$, 
h) $Q[m_p,n_{\bar{\Lambda}},l_\Lambda]$, 
i) $Q[m_p,l_{\bar{\Lambda}},n_\Lambda]$, 
j) $Q[n_p,m_{\bar{\Lambda}},m_\Lambda]$, 
k) $Q[n_p,m_{\bar{\Lambda}},l_\Lambda]$, and 
l) $Q[n_p,l_{\bar{\Lambda}},m_\Lambda]$
}
\end{figure*}

The remaining 12 directly measurable spin-correlations are shown in 
Fig.~\ref{others}
The first of these, Q[n$_p$], is
the analyzing power, often denoted A$_n$.  The remainder are
correlations between initial proton spin and components of one or both
final-state spins.  Errors are seen to be small
enough, even on most 3-spin
correlations, to allow structures to be clearly resolved.  This underscores
the advantage of fitting the spin-scattering matrix parameters.  If these 
directly measurable spin-correlations had been determined by a direct 
fit of Eq.~(\ref{eqn_b}) to the observed distribution, their errors would 
have been much larger and so meaningful structure would have been 
impossible to extract in most cases.

\begin{figure*}
\includegraphics[width=17.6 cm,clip]{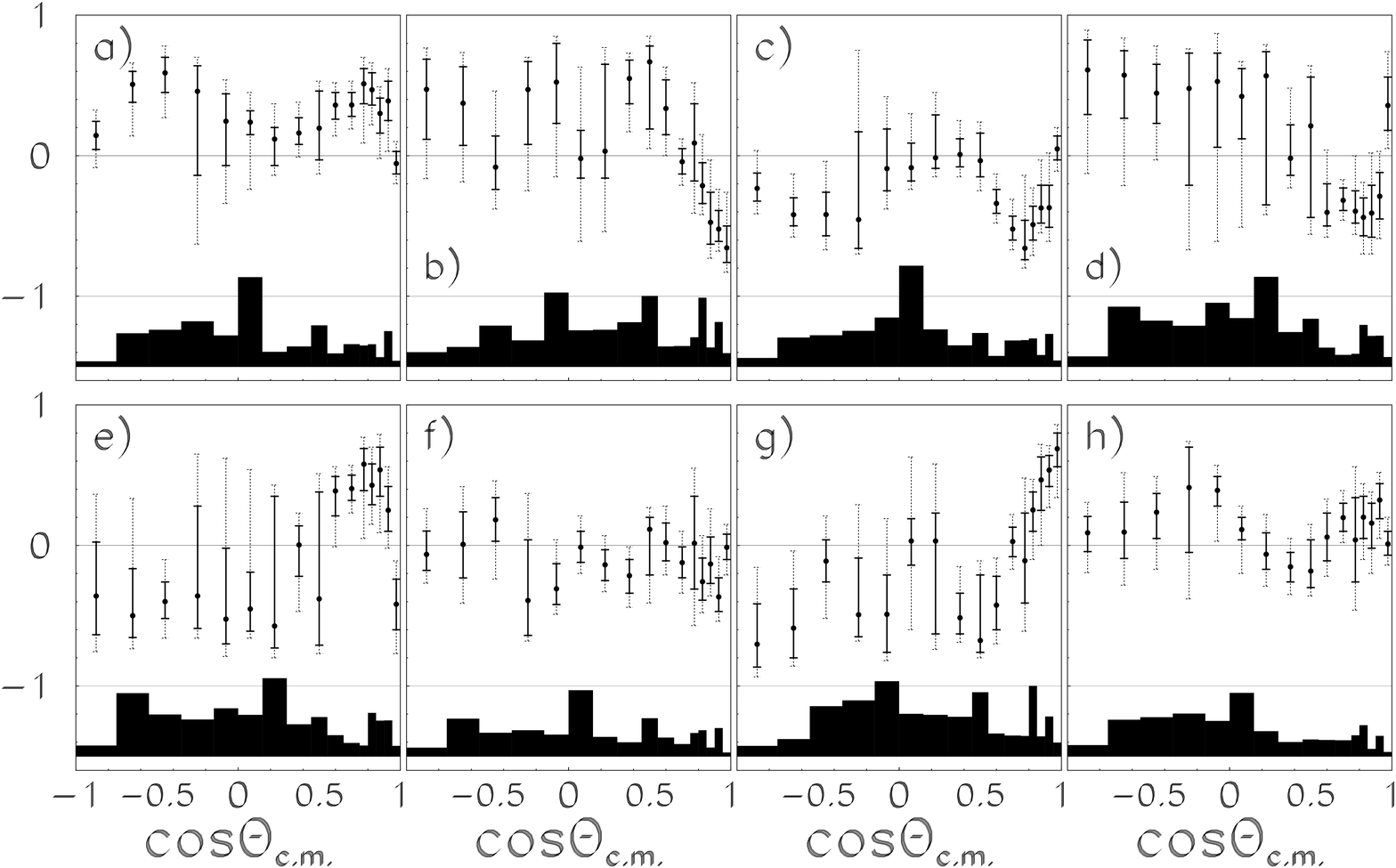}
    \caption{\label{longp}
Results for eight spin observables which do not
appear directly in the measured angular distribution.
Direct measurement of an individual one of these observables would
require a longitudinally polarized target proton but would not require
beam polarization. 
Statistical and systematic error estimates are displayed, as above. 
The spin observables displayed are 
a) $Q[l_p,m_\Lambda]$, 
b) $Q[l_p,l_\Lambda]$, 
 c) $Q[l_p,m_{\bar{\Lambda}}]$, 
 d) $Q[l_p,m_{\bar{\Lambda}},n_\Lambda]$, 
 e) $Q[l_p,n_{\bar{\Lambda}},m_\Lambda]$, 
 f) $Q[l_p,n_{\bar{\Lambda}},l_\Lambda]$, 
 g) $Q[l_p,l_{\bar{\Lambda}}]$, and
h)  $Q[l_p,l_{\bar{\Lambda}},n_\Lambda]$. 
}
\end{figure*}

\begin{figure*}
\includegraphics[width=17.6 cm,clip]{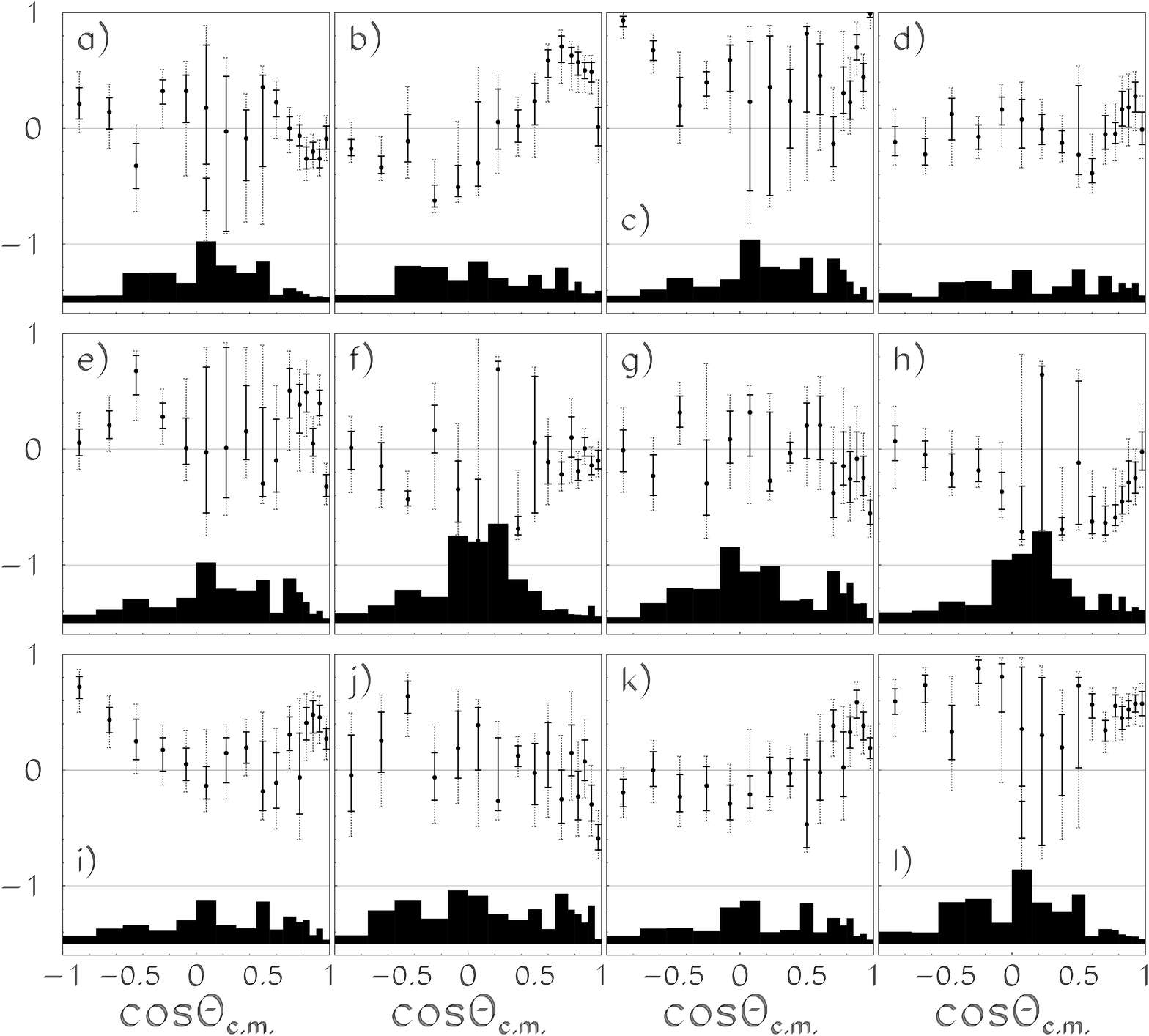}
    \caption{\label{polpbar}
Results for the final twelve spin observables which do not
appear directly in the measured angular distribution. 
Direct measurement of an individual one of these observables would
require a polarized anti-proton beam. 
Statistical and systematic error estimates are displayed, as
above. For one angular bin in each of
figures a) and l) double-minima have resulted in disjoint regions
falling within the the 1$\sigma$ limit.
The spin observables displayed are 
a) $Q[m_{\bar{p}},m_p]$, 
  b) $Q[m_{\bar{p}},m_p,n_\Lambda]$, 
 c) $Q[m_{\bar{p}},m_p,m_{\bar{\Lambda}},m_\Lambda]$, 
 d) $Q[m_{\bar{p}},m_p,m_{\bar{\Lambda}},l_\Lambda]$, 
 e) $Q[m_{\bar{p}},m_p,l_{\bar{\Lambda}},l_\Lambda]$, 
 f) $Q[m_{\bar{p}},l_p]$, 
 g) $Q[m_{\bar{p}},l_p,n_\Lambda]$, 
 h) $Q[m_{\bar{p}},l_p,m_{\bar{\Lambda}},m_\Lambda]$, 
i) $Q[m_{\bar{p}},l_p,m_{\bar{\Lambda}},l_\Lambda]$, 
j) $Q[m_{\bar{p}},l_p,n_{\bar{\Lambda}}]$, 
 k) $Q[m_{\bar{p}},l_p,l_{\bar{\Lambda}},m_\Lambda]$, and 
 l) $Q[l_{\bar{p}},l_p]$. 
}
  
\end{figure*}

Of the 256 spin-correlations,
$Q[j_{\bar{p}},k_p,\mu_{\bar{\Lambda}},\nu_\Lambda]$, defined by 
Eq.~(\ref{eqnQ}), one is trivially unity and 128 are constrained to be zero by 
parity conservation of the strong interaction.  An additional 88 can be 
neglected because symmetry requires that they are identical to (or the 
negative of) another one which is being considered.  In this sense there are
39 non-trivial spin correlations in addition to $I_0$.  These cannot be 
said to be forty independent observables since they can be expressed in 
terms of just eleven real parameters of the spin-scattering matrix at 
each $\Theta_{c.m.}$.  However there are 40 quantities which would be 
directly measurable given arbitrary beam and target polarization.  Of
these, twenty (including $I_0$) are directly measurable in the present 
experiment and have been presented above.  However, since the spin-scattering 
matrix is fully determined, the remaining 20 spin-correlations can equally 
well be extracted just as the directly measurable ones are in the present
analysis.  The eight such observables shown in Fig.~\ref{longp} would
be directly measurable (i.e.~would appear explicitly in the
description of the angular distribution) without a polarized beam 
if the target were
longitudinally polarized.  Although the target polarization in the
present experiment is purely transverse, these observables are still
determined, and in some cases determined quite accurately, in this
present measurement.  Similarly, Fig.~\ref{polpbar} gives the results for
spin observables which would be directly measurable only if the
antiproton beam were polarized.  Again, some of these are quite well
determined by the present data set.  In rare cases a double-minimum in
the log-likelihood function, ${\cal M}$, results in disjoint regions
falling within the 1$\sigma$ limit.  These are indicated in
Fig.~\ref{polpbar} by a second disjoint error bar.

\section{Discussion and Conclusions}

The method of determining the spin-scattering matrix, suggested in
\cite{Kent_Brian} has been successfully applied in practice.  
This is a unique case in which the full spin structure of a
 two-fermion interaction has been determined from a single
 measurement.  The self-analyzing property of hyperons combined with a
 transversely polarized target allows this unusual access to the spin
 structure of the production of strange-anti-strange quark pairs.
 The data set of about 2000 events per $\Theta_{c.m.}$-bin has proven
sufficient to accurately determine the parameters and from them to
learn the spin-correlations as well as other functions such as $S_F$
and $<\frac{d\sigma}{d\Omega}>$.  By construction, the results are
guaranteed to be internally consistent, obeying all constraints imposed
by the symmetries of the strong interaction under parity and charge 
conjugation.  Numerical values for these results are available \cite{EPAPS}.

Apart from the small background subtraction at back-angles, all results shown
for each angular bin have been obtained completely independently of the results
at other angular bins and are based on non-overlapping sets of events.  The 
smooth variations as a function of $\Theta_{c.m.}$
seen in most of the spin-correlations is in no way built-in to the analysis
technique.  The fact that the angular variation of the data appears smooth
 is a reassurance that the entire chain of event reconstruction
and data analysis is performing reasonably and extracting meaningful results.
Similarly the fact that the bin-to-bin 'scatter' in the data appears to be 
consistent with
the assigned error bars is an independent verification of the validity of the 
method of error analysis.

The most important aspects of these results, relating to 
$Q[n_p,n_\Lambda]$ and $Q[n_p,n_{\lbar}]$ have already been presented
in \cite{Bernd_et_al}.  As shown in Fig.~\ref{Kent_8.2}, the
measured values of these spin-correlations differ
markedly from the predictions of a meson exchange (MEX) model
\cite{MEX_pred} (solid line) and a quark-gluon-inspired (QG)
model \cite{QG_pred} (dotted line) despite the
fact that both these models reasonably describe the significant 
spin-structure observable with an unpolarized target.

All MEX models generally predict a large tensor interaction which couples
spin-triplet $\pbarp$ initial state to spin-triplet $\lbarl$ final
state, flipping the spin in the process.  For this reason, both the
spin transfer, $Q[n_p,n_{\lbar}]$, and the depolarization,
$Q[n_p,n_\Lambda]$, are predicted to be strongly negative. This
prediction holds even in the presence of initial- and final-state
interactions, which have been included in the prediction shown by
solid lines in the figures.  The measured values are far less strongly
negative than the predictions and, in fact, are positive at forward 
angle indicating that both final-state particles tend to be
aligned with the initial proton spin.  Furthermore, $Q[n_p,n_{\lbar}]$
is strongly positive at back angles meaning that the normal component
of the proton's spin is transferred to the $\lbar$ in contrast to
the MEX prediction.

All existing calculations using QG models have been 
restricted to $^3P_0$, 'vacuum' terms,
and $^3S_1$ 'gluon' terms.  So the interaction is purely spin triplet,
having $S_F=0$ built into the model.  Here there is a much smaller
tensor interaction and so less spin-flip.  It was because of this
characteristic difference between QG and MEX models that
$Q[n_p,n_\Lambda]$ was first suggested \cite{Holinde} as an
observable which would distinguish experimentally between the two
classes of model.  A vanishing singlet fraction necessarily implies
that $Q[n_p,n_\Lambda] = Q[n_p,n_{\lbar}]$.  This is reflected in the
predictions of the QG model shown as dotted lines in Fig.~\ref{Kent_8.2}.
As shown in Fig.~\ref{Kent_7.23}, however, the singlet fraction is
distinctly non-zero at back angles, in violation of the assumptions of
the existing models.  This manifests itself in the data as a large
difference in back-angle behavior of the two distributions.

\begin{figure*}
\includegraphics[width=17.6 cm,clip]{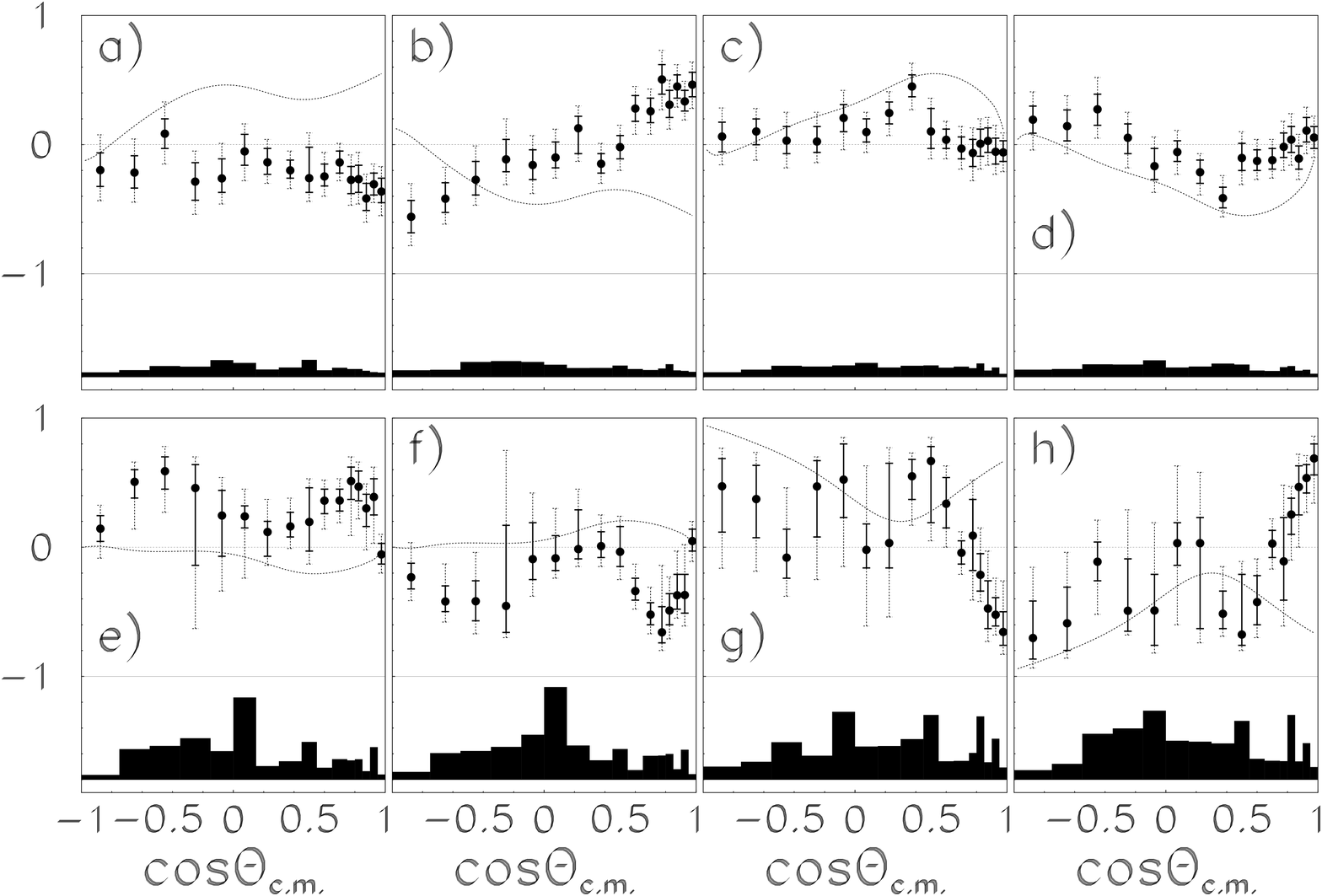}
    \caption{\label{Kent_8.4}
Comparison of quark-gluon predictions \protect\cite{QG2} to
    measured spin-transfer and depolarization observables.
    The measured results have been presented in 
Figs.~\ref{others} and \ref{longp}
and are repeated here for
    comparison with the predictions.
Statistical and systematic
error estimates on the data are displayed, as above. 
The spin observables displayed are 
a) $Q[m_p,m_\Lambda]$,
b) $Q[m_p,m_{\bar{\Lambda}}]$,
c) $Q[m_p,l_\Lambda]$,
d) $Q[m_p,l_{\bar{\Lambda}}]$,
e) $Q[l_p,m_\Lambda]$,
f) $Q[l_p,m_{\bar{\Lambda}}]$,
g) $Q[l_p,l_\Lambda]$, and 
h) $Q[l_p,l_{\bar{\Lambda}}]$.
}  

\end{figure*}

Apart from large $\Theta_{c.m.}$ and very small $\Theta_{c.m.}$, the QG
model is seen to dramatically over-predict both  $Q[n_p,n_\Lambda]$
and $Q[n_p,n_{\lbar}]$.  Predictions have also been made \cite{QG2} for the
transfer of other components of proton spin to the spin of the
$\Lambda$ and $\lbar$.  These are compared to the data in 
Fig.~\ref{Kent_8.4}. While the disagreement is not as striking, partly
because of the relatively larger error bars, it is clear that
significant modification of the model will be needed to match these
correlations and to predict the others reported here.  A modification
which is clearly required is inclusion of singlet strength, such as
$^1S_0$.

While a great wealth of information has been gained on the spin
dynamics of $\reac$ at $p_{\pbar} = 1.637$ GeV/c,  few
conclusions can be drawn because theoretical models lag
significantly behind in understanding the data at this point.
Availability of this data may inspire increased theoretical activity.

Among the most precisely determined spin correlations is one which has
not previously been measured.  The analyzing power (a correlation of
spin only with scattering angle, not with other spins) $Q[n_p]$ 
(Fig.~\ref{others}a), usually
denoted $A_n$, is constrained to vanish at $\cos(\Theta_{c.m.})=\pm 1$ 
but has now been determined to be strongly positive in the forward hemisphere
and mostly negative for back angles.  The complex angular structures
of the 2-spin correlations, 
$Q[m_{\bar{\Lambda}},m_\Lambda]$, 
$Q[m_{\bar{\Lambda}},l_\Lambda]$, 
$Q[n_{\bar{\Lambda}},n_\Lambda]$,
and 
$Q[l_{\bar{\Lambda}},l_\Lambda]$, (Fig.~\ref{Kent_7.22} a through d,
respectively) 
although consistent with previous measurements \cite{Horsti}, is now
revealed in far more detail.  Among the 3-spin correlations, some of
the directly measurable ones show the cleanest structure with 
$Q[n_p,m_{\bar{\Lambda}},l_\Lambda]$ (Fig.~\ref{others}k) and
$Q[n_p,l_{\bar{\Lambda}},m_\Lambda]$ (Fig.~\ref{others}l) 
showing remarkably similar
behavior, while it is not trivially expected that they should be
equal.  A distinctly different
structure is seen in $Q[n_p,m_{\bar{\Lambda}},m_\Lambda]$ 
($=-Q[n_p,l_{\bar{\Lambda}},l_\Lambda]$) (Fig.~\ref{others}j) 
which is more compressed to
forward angles.  Even some of the 4-spin correlations are determined
well enough to unveil distinct angular structure. Regions of
non-vanishing correlation are seen for example in
$Q[m_{\bar{p}},l_p,m_{\bar{\Lambda}},m_\Lambda]$,
$Q[m_{\bar{p}},l_p,m_{\bar{\Lambda}},l_\Lambda]$, and
$Q[m_{\bar{p}},l_p,l_{\bar{\Lambda}},m_\Lambda]$ (Figs.~\ref{polpbar}h,
\ref{polpbar}i, and \ref{polpbar}k, respectively). 
It would be fruitless to speculate on the meaning of each of these
structures individually.  A coherent picture will required theoretical
modeling to simultaneously explain all available spin-correlations or
equivalently to directly predict the coefficients of the
spin-scattering matrix.  This difficult task is made all the more
difficult by the strength of the initial- and especially final-state
interactions.  But even in their presence, the physics reduces to just
eleven real parameters at each angle.  There may be advantages to comparing
theories to the experimentally determined spin-matrix parameters.  They 
may tie in more 
directly to the underlying spin physics of the model.
Also the process of determining the quality of the agreement is
simplified by removing the 'double-counting' which is inherent in
comparing up to 40 spectra when all the physics reduces to just eleven
sets of parameters.

\begin{acknowledgments}
The members of the PS185 collaboration thank the LEAR/CERN accelerator team.
We also gratefully 
acknowledge financial and material support from the German Bundesministerium 
f\"{u}r Bildung und Forschung, the Swedish Natural Science Research Council, 
the United States Department of Energy under contracts DE-FG02-87ER40315
and DE-FG03-94ER40821, and the United States National Science Foundation. 
\end{acknowledgments}


\end{document}